\documentclass[showpacs,pre,amssymb,amsmath,preprint]{revtex4}

\usepackage{bm}
\usepackage{graphics}
\usepackage{multirow}		
\usepackage{rotating}		
\numberwithin{equation}{section}

\usepackage{textcomp}		
\usepackage{psfrag}
\usepackage{amsmath}
\usepackage{amssymb}
\usepackage{textcomp}		
\usepackage{array}		
\usepackage{multirow}		
\usepackage{rotating}		

\def \dd{\hbox{d}}

\def \dd{\hbox{d}}

\newcommand{\be}{\begin{equation}}
\newcommand{\ee}{\end{equation}}
\newcommand{\bea}{\begin{eqnarray}}
\newcommand{\eea}{\end{eqnarray}}
\def \la{\label}

\def\({\left (}
\def\){\right )}
\def\]{\right]}
\def\[{\left[}
\def\<{\left <}
\def\>{\right>}
\newcommand{\bX}{\mathbf{X}}

\newcommand{\br}{\mathbf{r}}
\newcommand{\bk}{\mathbf{k}}

\newcommand{\bp}{\mathbf{p}}

\newcommand{\cV}{\mathcal{V}}

\newcommand{\cC}{\mathcal{C}}

\newcommand{\bR}{\mathbf{R}}
\newcommand{\cL}{\mathcal{L}}

\newcommand{\cU}{\mathcal{U}}

\newcommand{\na}{\pmb{\nabla}}

\begin{document}

\title{Quantum Coulomb systems : screening, recombination and van der Waals forces}
\author{A.Alastuey}
\affiliation{Laboratoire de Physique, ENS Lyon and CNRS, 
46 all\'ee d'Italie, 69364 Lyon Cedex 07, France}
\date{\today}

\begin{abstract}

The study of quantum Coulomb systems at equilibrium 
is important for understanding properties of 
matter in many physical situations. Screening, recombination and 
van der Waals forces are basic phenomena which result from the 
interplay of Coulomb interactions, collective 
effects and quantum mechanics. Those phenomena are introduced in
the first part of this lecture, through various 
physical examples. Their treatment within mean-field theories 
and phenomenological approaches is also exposed, while 
related predictions are discussed.
This sheds light on fundamental issues, which must be analyzed 
without any \textsl{a priori} approximations or modelizations. 
The second part of this lecture is precisely devoted to 
the presentation of various exact results for the 
quantum proton-electron hydrogen plasma. Such results are derived 
within the Screened Cluster Representation, which is constructed by  
combining the path integral representation of 
the Coulomb gas with Mayer-like diagrammatical techniques. 
They illustrate the breakdown of Debye exponential screening 
by quantum fluctuations, as well as the emergence of familiar chemical 
species in suitable low-temperature and low-density limits. Also, the
amplitude of van der Waals forces is shown to be reduced by free charges.

\end{abstract}

\maketitle

\section{Introduction}

Under standard Earth conditions, and also in many astrophysical situations, the 
properties of matter result from the interplay between non-relativistic quantum 
mechanics and Coulomb interactions. All relativistic effects, as well as the other 
non-electromagnetic interactions can be safely omitted. A description of matter 
in terms of quantum nuclei and quantum electrons interacting \textsl{via} the 
Coulomb potential is then sufficient. In that context, the derivation of exact results for
equilibrium properties of quantum Coulomb systems is of crucial 
importance.

\bigskip

The study of quantum Coulomb systems within statistical mechanics, 
requires to face several difficult problems related to both short- and 
long-range specificities of Coulomb potential, as well as to quantum 
mechanics itself, namely screening and recombination. 
In the first introductory part of the present lecture, 
those problems are successively adressed for various 
physical examples where free charges interacting \textsl{via} the
Coulomb potential are present. First, it is instructive 
to consider screening in classical systems where quantum effects 
can be omitted. According to the pioneering 
Debye-like mean-field theories, Coulomb interactions 
are exponentially screened at the classical level, 
a prediction confirmed by a large variety 
of rigorous proofs and exact results. In the 
quantum case, similar mean-field theories also predict 
an exponential decay of equilibrium particle correlations. 
However, they do not account for a very fundamental feature 
of quantum mechanics, namely the intrusion of dynamical effects 
at equilibrium. Hence, there exist various arguments which 
suggest a breakdown of exponential screening in quantum 
systems. In the literature, recombination has been mainly 
dealt with in the framework of the chemical picture. The 
corresponding phenomenological 
approaches are based on \textsl{ad hoc} modelizations for 
preformed entities and their interactions. 

\bigskip

As illustrated by the various considerations exposed in 
the introductory part of this lecture, screening, recombination 
and van der Waals forces result from entangled mechanisms 
combining Coulomb interactions and quantum mechanics. Although, 
phenomenological treatments of such phenomena have 
been widely developped, there still remain 
subtle effects, a deeper understanding of which requires a more 
fundamental analysis in the framework of 
quantum Coulomb systems, namely the 
so-called physical picture. The aim of the second part of this lecture is 
to present several exact results relative to fundamental issues 
about screening and recombination. For the sake of simplicity and pedagogy, 
we consider the quantum hydrogen plasma made with point protons and 
point electrons interacting \textsl{via} the Coulomb potential. First, 
we present rigorous proofs about thermodynamical stability 
on the one hand, and the atomic limit on the other hand. Then, 
we introduce the Feynman-Kac path integral representation, 
which turns out to be a quite efficient tool for our purpose. 
Its application to a many-body quantum system  provides an equivalent 
classical system made with extended objects called loops. Standard 
Mayer-like diagrammatical series for the gas of loops, 
are exactly transformed into the Screened Cluster Representation for 
equilibrium quantities of the quantum system. That transformation, 
based on suitable resummations and reorganizations, accounts 
simultaneously for both screening and recombination. Quantum fluctuations, 
which play a crucial role in screening, are merely embedded in 
loop shapes. Recombination, which cannot be treated perturbatively 
in the Coulomb potential, is automatically ensured by the presence of 
Boltzmann factors associated with loop interactions. The
Screened Cluster Representation is applied to hydrogen in the Saha regime, 
defined by both low temperatures and low densities, 
where it behaves as a partially ionized atomic gas. An exact 
asymptotic expansion for the equation of state is constructed, 
beyond familiar Saha theory. It sheds light on 
a suitable first-principles account of contributions of usual 
chemical species, without any \textsl{a priori} modelizations.
Also, particle correlations are found to decay algebraically. Thus, 
ionized protons and ionized electrons only reduce the amplitude 
of van der Waals interactions between atoms. As a conclusion, we 
summarize the main answers to the above issues inspired by those 
exact results.

\section{Coulomb interactions and quantum mechanics at work}

\subsection{Examples and specific features} \label{ssexamples}

A description of matter in terms of Coulomb systems, 
also called plasmas, is 
necessary as soon as free charges are present. There is 
a large variety of physical examples where such  
situations occur. For instance, electrolytes involve ionic 
species obtained by dissolution of salts into water. Also, in metals, the 
electrons of the conduction band freely move across the 
samples. Other examples can be found in astrophysical 
situations, where high temperatures or high pressures 
favor ionization of matter, in general inside the cores of 
compact objects or stars.

\bigskip

Let us consider a plasma made with point charges. 
Two charges $q_i$ and $q_j$ located at $\br_i$ and $\br_j$ interact \textsl{via} 
the instantaneous Coulomb potential 
\be
\la{coulombpot}
u_{{\text{C}}}(\br_i,\br_j)= \frac{q_i q_j}{|\br_i-\br_j|} \;,
\ee
while the full interaction potential of the system reduces to the 
sum of pairwise interactions (\ref{coulombpot}).
Here, we dot not taken into account 
retardation effects or magnetic forces, 
and we discard the coupling of charges to elecromagnetic radiation.  
That purely Coulombic description is sufficient as far as 
the average speed of charges is small compared to the speed of light $c$. 
Roughly speaking, for a classical charge with mass $m$, this requires that the
thermal energy $k_BT$ is small compared to the rest energy $mc^2$
\footnote{When quantum effects 
become important, $k_BT$ is no longer the relevant energy which 
determines the average speed of the considered charge. For instance, conduction 
electrons in metals are highly degenerate and we have to compare the Fermi 
energy to $mc^2$.}. 
This implies $T \ll 10^{10} K$ for electrons 
and  $T \ll 10^{13} K$ for protons. Such conditions are 
fullfilled in many physical systems!

\bigskip

Contrarily to interactions between neutral entities, 
Coulomb potential is long ranged, namely it is not 
integrable at large distances 
\be
\la{longrange}
\int_{r>D} \dd \br \, \frac{1}{r} = \infty \; ,
\ee
where $D$ is some irrelevant cut-off length.
That divergence might pollute thermodynamical quantities,  
and usual extensivity properties might be lost. The underlying 
physical picture is that 
charges with the same sign tend to repell together 
too strongly at large distances, 
so the system might explode. 

\bigskip

In addition to above long-range problems, the $1/r$-nature of 
the Coulomb potential causes also some trouble at short 
distances. If we consider two classical opposite point charges $\pm q$
separated by a distance $r$, the 
corresponding Boltzmann factor is not integrable at $r=0$, 
\textsl{i.e.}
\be
\la{shortrange}
\int_{r<D} \dd \br \, \exp \( \frac{\beta q^2}{r} \) = \infty 
\ee
with $\beta=1/(k_BT)$. Thus, a classical system with positive 
and negative point charges collapses.

\bigskip

According to above considerations, the study 
of Coulomb systems in the framework of statistical mechanics 
has to face two central difficulties related to the behaviours of 
the Coulomb potential at respectively
large ($r \rightarrow \infty$) and short ($r \rightarrow 0$) distances.
In the following, we will describe the fundamental mechanisms 
which cure the corresponding singularities, namely screening and 
recombination. Screening is a collective effect which 
prevents explosion at large distances. Recombination is 
a purely quantum mechanical effect, where the 
uncertainty principle smears out the divergency of 
the Coulomb potential at $r=0$, so collapse is avoided. 
Those mechanisms are adressed successively, by 
considering physical systems for which simplified models 
and/or phenomenological approaches can be introduced.

\bigskip

First, in Section~\ref{ssdebye}, we present Debye theory for 
a classical electrolyte. That mean-field approach predicts 
an exponential screening of Coulomb interactions, which is 
confirmed by rigorous results. A similar mean-field theory 
has been applied to the quantum electron gas, as 
described in Section~\ref{ssquantumscreening}. Again, an 
exponential screening is found, but that prediction 
is shown to be quite doubtful according to the unavoidable intrusion 
of dynamical effects in equilibrium quantities of 
quantum systems. Recombination and van der Waals 
forces are considered in Section~\ref{ssrecombination} through the 
example of hydrogen in the Sun, the state of which changes from a fully ionized 
plasma in the core to an atomic gas in the photosphere. We present 
simple arguments, as well as some standard phenomenological considerations, 
which introduce the more sophisticated analysis presented in the 
second part of this lecture.

\subsection{Screening in classical systems} \label{ssdebye}

\subsubsection{Primitive model for an electrolyte}

Let us consider an ordinary solution of 
sodium chloride. For a concentration 
$C_0=0.1$ moles/l at room temperature $T=300$ K, 
the salt is entirely dissociated, so the solution 
reduces to a mixture of ions $\text{Na}^+$, $\text{Cl}^-$
and water molecules $\text{H}_2\text{O}$ (ions $\text{H}_3\text{O}^+$ 
and $\text{OH}^-$  
are omitted since their concentration is negligible 
compared to $C_0$). The mean interionic distance,  
$a=(3/(4\pi\rho))^{1/3}$ with the common
ionic number densities $\rho=\rho_{+}=\rho_{-}$, is large 
compared to both the typical size $d$ of ions and 
the mean distance between water molecules. Therefore, 
a large number of water molecules surrounds each ion, 
and water may be reasonably replaced by a
continuous medium with dielectric constant 
$\epsilon_{{\text{w}}}$. Also the 
de Broglie thermal wavelengths 
$\lambda_{\pm}=(\beta \hbar^2/M_{\pm})^{1/2}$
are small compared to $d$, so ions can be treated 
classically. 

\bigskip

According to the above hierarchy of length scales, the 
present electrolytic solution is well described by a 
classical two-component plasma made with hard spheres 
carrying charges $q_{\pm}=\pm e$, also called the 
restrictive primitive model in the literature. 
Two ions of species 
$\alpha=\pm$ and $\gamma=\pm$ separated by a distance $r$ 
interact \textsl{via} the potential
\bea
\la{ionicpot}
u_{{\alpha\gamma}}(r) & = &+\infty \; ,\; r < d_{{\alpha\gamma}} \nonumber \\
u_{{\alpha\gamma}}(r) & = & \frac{q_{{\alpha}} q_{{\gamma}} }
{\epsilon_{{\text{w}}} r} \; ,\; r > d_{{\alpha\gamma}} \; .
\eea
The hard-core part of the potential is a phenomenological 
description of the short-range repulsion between the electronic clouds of 
the two considered ions. Of course, within that modelization, no collapse 
between oppositely charged ions occur. Also, notice that the bare Coulomb 
potential is renormalized \textsl{via} the 
standard factor $1/\epsilon_{{\text{w}}}$ which accounts for the 
underlying presence of a continuous dielectric medium.

\bigskip

In order to adress the question of screening, we 
introduce the equilibrium charge density $C_{{\alpha}}(\br)$ 
surrounding an ion with species $\alpha$ fixed 
at the origin, which reads
\be
\la{chargedensity}
C_{{\alpha}}(\br) = \sum_{\gamma} q_{{\gamma}} 
\frac{\rho_{{\alpha\gamma}}^{(2)}(\mathbf{0},\br)}{\rho_{{\alpha}}} \,
\ee
where $\rho_{{\alpha\gamma}}^{(2)}$ is the two-point equilibrium distribution 
of species $\alpha$ and $\gamma$. Screening can be easily 
understood within a mean-field calculation of $C_{{\alpha}}(\br)$.
The basic ideas sustaining mean-field approach have been first
introduced by Gouy~\cite{Gou} and Chapman~\cite{Cha} 
fot the study of electrical double layers near charged electrodes. 
They have been extended to the calculation of polarization clouds 
in the bulk phase by 
Debye and Huckel~\cite{DebHuc}. Here, we present the corresponding 
arguments which provide the mean-field form of $C_{{\alpha}}(\br)$.

\subsubsection{Debye theory}
\begin{figure}
\psfragscanon
\psfrag{b0}[c][c]{$\mathbf{0}$}
\psfrag{br}[c][c]{$\mathbf{r}$}
\psfrag{Na}[c][c]{Na$^+$}
\psfrag{Cl}[c][c]{Cl$^-$}
\includegraphics[width=0.6\textwidth]{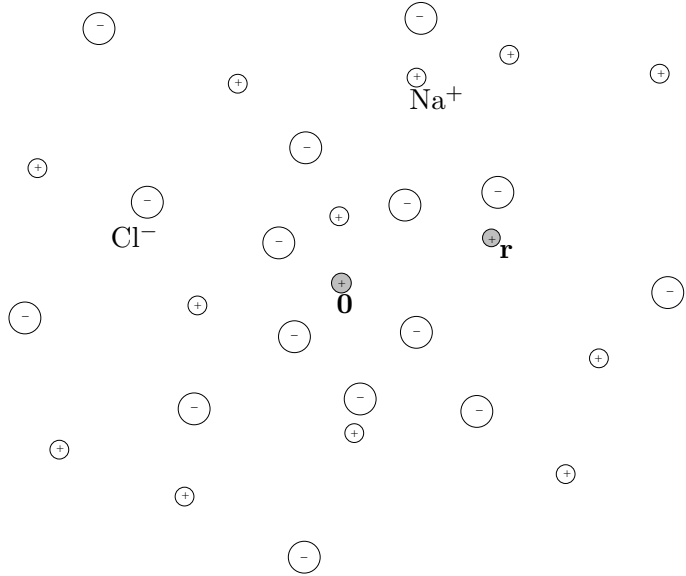}
\caption{\label{NaCl}A typical ionic configuration $\cC$ where two cations 
$\text{Na}^+$ ($\alpha=\gamma=+$) are fixed at 
$\mathbf{0}$ and $\br$ respectively.}
\end{figure}
Let $\cC$ be a given spatial configuration of ions where one ion 
$\alpha$ is fixed at the origin and one ion $\gamma$ is 
fixed at $\br$. In Fig.~\ref{NaCl}, we draw a typical configuration 
which illustrates that ion $\alpha$ attracts  
oppositely charged ions and repells the other ones. Ion $\gamma$ 
feels the electrostatic potential $\varphi(\br|\cC)$ created at $\br$ 
by all the remaining ions, in particular those which are far 
apart, because the Coulomb potential is long ranged. Since 
$\varphi(\br|\cC)$ is a sum of a large number of terms,
fluctuations of the positions of non-fixed 
ions should slightly modify its value. That statement is inspired 
by the law of large numbers, and it can been applied 
to other many-body problems where long range interactions are 
present~\footnote{See for instance a one-dimensional classical gas where 
the attractive part of the pair-potential has an infinite range and a
vanishingly small amplitude~\cite{Kac}, or lattice systems described by 
Ising Hamiltonians where any two spins interact with a 
constant and vanishingly small coupling (see e.g. Ref.~\cite{Bar}). In those double 
zero-coupling and infinite-range limits, the mean-field treatment becomes exact.}. 
Thus, it is 
tempting to replace $\varphi(\br|\cC)$ by its average value 
$\varphi_{\alpha}(\br)$, which is nothing but the electrostatic 
potential created by ion $\alpha$ plus its polarization cloud with
charge distribution $C_{{\alpha}}(\br)$.
Within that mean-field treatment, the density 
of ions $\gamma$ at $\br$ is merely given by the Boltzmann 
distribution of an ideal gas in the external potential 
$q_{{\gamma}} \varphi_{\alpha}(\br)$, 
\textsl{i.e.}
\be
\la{boltzmann}
\frac{\rho_{{\alpha\gamma}}^{(2)}(\mathbf{0},\br)}{\rho_{{\alpha}}} = 
\rho_{{\gamma}} \exp (-\beta q_{{\gamma}} \varphi_{\alpha}(\br)) \,.
\ee

\bigskip

At this level, we have completely omitted short-range effects 
arising from the hard-core part of ion-ion potential (\ref{ionicpot}). 
Under diluted conditions for which $d \ll a$, this is legitimate since for 
most configurations $\cC$, ion $\gamma$ only feels the Coulomb parts of 
its interactions with other ions. At the same time, 
position fluctuations of neighbours of fixed ion $\gamma$ 
can be neglected, only if they
generate variations of $\varphi(\br|\cC)$ which do not exceed the 
thermal energy $k_BT$. This implies that the 
corresponding typical value $e^2/(\epsilon_{{\text{w}}} a)$ of Coulomb interactions 
is small compared to $k_BT$, a condition indeed fulfilled at 
low densities. According to previous considerations, the linearization of 
the Boltzmann factor (\ref{boltzmann}) is consistent with the mean-field 
treatment, and it provides 
\be
\la{boltzmannlin}
C_{{\alpha}}(\br) = -\beta  \sum_{\gamma} q_{{\gamma}}^2 \rho_{{\gamma}} 
\varphi_{\alpha}(\br) \,, 
\ee
where we have used the neutrality condition 
\be
\la{neutrality} 
\sum_{\gamma} q_{{\gamma}} \rho_{{\gamma}} = 0 \, .
\ee
The insertion of the linearized expression 
(\ref{boltzmannlin}) of $C_{{\alpha}}(\br)$ into 
Poisson equation for $\varphi_{\alpha}(\br)$, 
shows that $\varphi_{\alpha}(\br)=q_{{\alpha}}\phi_{{\text{D}}}(\br)$, where $\phi_{{\text{D}}}$ is the solution of 
\be
\la{helmoltz}
\(- \Delta + \kappa_{{\text{D}}}^2 \) \phi_{{\text{D}}}(\br)=
\frac {4 \pi}{\epsilon_{{\text{w}}}} \; \delta(\br)
\ee 
with 
\be
\la{kappa}
\kappa_{{\text{D}}}^2= \frac {4 \pi}{\epsilon_{{\text{w}}}} 
\beta \sum_{\gamma} q_{{\gamma}}^2 \rho_{{\gamma}}=
\frac {8 \pi}{\epsilon_{{\text{w}}}}\beta e^2\rho
\ee
and boundary conditions 
\be
\la{boundary}
\phi_{{\text{D}}}(\br) \rightarrow 0 \;\;\;\; \text{when} \;\;\;\; r \rightarrow \infty \; .
\ee 
Such boundary conditions amount to impose that the ionic densities 
induced at $\br$ tend to the bulk homogeneous value $\rho$ 
when $r \rightarrow \infty$, or in other words to 
make the quite plausible assumption that polarization effects decay 
at large distances. 

\bigskip

The Helmholtz equation (\ref{helmoltz}) together with boundary conditions 
is easily solved in Fourier space, and 
$\phi_{{\text{D}}}(\br)$ is then readily obtained by applying 
Cauchy's theorem (see \textsl{e.g.} Ref.~\cite{AlaMagPuj} for 
a detailed presentation of that standard calculation), 
\be
\la{debyepot}
\phi_{{\text{D}}}(\br)=\phi_{{\text{D}}}(r)= 
\frac {\exp (-\kappa_{{\text{D}}} r)}{\epsilon_{{\text{w}}}r} \; .
\ee
Thus, the electrostatic potential 
$q_{{\alpha}}\phi_{{\text{D}}}(r)$ 
inside the electrolytic solution decays exponentially faster 
than the potential $q_{{\alpha}}/(\epsilon_{{\text{w}}}r)$ 
created by a single ion $\alpha$ immersed in water. In the electrolyte, 
such an ion is screened by its induced polarization cloud over 
the so-called Debye length $\lambda_{{\text{D}}}=\kappa_{{\text{D}}}^{-1}$. 

\subsubsection{Reliability and limits of mean-field predictions}

It is instructive to determine what are the physical conditions 
under which the mean-field treatment is reliable at a 
quantitative level. The main argument for neglecting 
fluctuations relies on the large value of the number of ions which 
contribute to the electrostatic potential at $\br$. 
In average, those ions are contained in a 
sphere with radius $r$ as a consequence of Gauss theorem, so their
number is of order $8\pi\rho r^3/3$. That number must be large 
for $r$ of order the typical length 
scale $\lambda_{{\text{D}}}$. This implies 
$a\kappa_{{\text{D}}} \ll 1$, or equivalently the weak-coupling condition 
$e^2/(\epsilon_{{\text{w}}} a k_BT) \ll 1$. Within 
that condition, the linearization of Boltzmann factor (\ref{boltzmann})
is also justified. If that condition is fulfilled at high temperatures or 
low densities, hard-core effects can be neglected only if $d \ll a$. 
The previous mean-field approach is then expected to be valid, at least 
at a quantitative level, at sufficiently high temperatures and low densities.
Notice that in the present example, the weak-coupling condition is 
ensured thanks to the large value of the water dielectric constant, 
$\epsilon_{{\text{w}}} \simeq 80$, which drastically reduces the strength 
of electrostatic interactions. As a consequence, Debye theory works 
reasonably well for most electrolytic solutions of ordinary salts 
under standard conditions.
 
\bigskip

Beyond its practical interest for a large variety of physical systems, 
Debye theory suggests a fundamental result, namely the 
exponential decay of correlations in classical charged systems, at 
least in fluid phases. In fact, that 
remarkable property has been proved for various systems \cite{BryFed,Imb,Ken}
at sufficiently high temperatures and sufficiently low densities, 
\textsl{i.e.} in thermodynamical regimes where 
mean-field approach is expected to work. The very difficult part of those 
proofs relies on the complete control of the contributions of 
all the effects omitted in Debye theory. 
It turns out that such effects do not destroy 
the exponential clustering predicted by mean-field. In other words, 
the essence of classical exponential screening is 
captured by mean-field approach, which enlights its 
collective nature. Also, the harmonicity of the Coulomb 
potential is a key ingredient, as illustrated by an analysis 
of the equilibrium BGY hierarchy~\cite{AlaMar1}.
In general, the fast decay of particle correlations 
is proved to be related to a perfect arrangment of polarization 
clouds, as exemplified through multipole sum rules~\cite{Mar1}. 
Here, the total charge of the polarization cloud 
surrounding $\alpha$ must exactly cancel out its charge, namely
\be
\la{charge}
\int \dd \br \; C_{{\alpha}}(\br) = - q_{{\alpha}} \; .
\ee
That monopole sum rule is indeed satisfied by the mean-field expression
\be
\la{meanfield}
-q_{{\alpha}}\kappa_{{\text{D}}}^2 \frac {\exp (-\kappa_{{\text{D}}} r)}{4\pi r}
\ee
of $C_{{\alpha}}(\br)$. Of course, the existence of positive and 
negative charges is crucial for screening of Coulomb interactions 
\footnote{In systems where particles interact \textsl{via} the 
gravitationnal $1/r$-potential, since all masses are positive, 
all particles attract each other so
no screening occurs~\cite{DauRuf,Pom}.}. The mean-field expression
(\ref{meanfield}) diverges when $r \rightarrow 0$. Such divergency 
is unphysical, and it proceeds from the linearization of 
Boltzmann factor (\ref{boltzmann}) which is never valid at short distances.
However, that spurious short-range singularity 
does not affect integrated quantities, like the total charge 
(\ref{charge}) or thermodynamical quantities.

\bigskip

At a quantitative level, corrections to Debye theory can be 
derived within Abe-Meeron diagrammatical expansions~\cite{Abe,Mee}. 
Debye theory is merely recovered by keeping only the first graph in those 
series. Each of the remaining graphs provides an exponentially 
decaying contribution to particle correlations, in agreement with 
above proofs. The resulting screening length $\lambda_{{\text{S}}}(\rho,T)$  
differs from its Debye expression $\lambda_{{\text{D}}}$, 
and it might be approximately estimated by selecting suitable classes of 
Abe-Meeron graphs. Notice that short-range repulsion contributes 
to $\lambda_{{\text{S}}}(\rho,T)$ in a rather subtle way, so a reliable 
description of the variations of $\lambda_{{\text{S}}}(\rho,T)$ when temperature 
is decreased and/or density is increased is a challenging problem. At sufficiently low 
temperatures and/or high densities, the occurence of phase transitions might 
lead to divergencies of $\lambda_{{\text{S}}}(\rho,T)$. For instance, exponential
screening might be lost in a crystalline phase, or at the critical point of the liquid-gas 
transition as suggested in Ref.~\cite{AquFis}. 

\bigskip

Eventually, let us conclude that section by a very important 
remark about dynamical aspects. In a real sample, 
ions continuously move, and because of their finite inertia, 
instantaneous perfect arrangments of the polarization clouds 
cannot be ensured. However, according to the somewhat magical 
recipe of statistical mechanics, at equilibrium the time-average 
of any quantity can be equivalently computed as a 
phase-space average with the Gibbs measure. Thanks to the 
remarkable factorization of that measure into 
the Maxwellian distribution of momenta times  
the Boltzmann factor associated with the interaction potential, 
dynamical features are washed out by 
momenta integration. Then, the calculation of equilibrium correlations
in position space becomes a purely static problem, where only 
configurations of particle positions must be considered 
without caring about particle velocities. In other words, 
dynamical fluctuations no longer intervene, so 
above considerations about particle inertia 
are not relevant. However, in a quantum system where above 
factorization of the Gibbs measure is no longer valid, 
we can anticipate that dynamical effects should 
contribute to equilibrium correlations.

\subsection{Screening in quantum systems} \label{ssquantumscreening}

\subsubsection{Jellium model for conduction electrons}

In ordinary metals, like Copper for instance, there exists a finite 
density $\rho$ of ionized electrons which ensure electrical conduction. 
Those electrons freely move among the remaining ions, 
which can be considered as fixed at their lattice sites. 
A further simplification is to replace the 
periodic ionic charge density by the constant $e\rho$. This 
provides a model for conduction electrons, called either jellium or 
one-component plasma, where such electrons are immersed in an uniform 
and rigid neutralizing background.

\bigskip

Under standard conditions and for simple metals with one or two 
ionized electrons per atom, the typical values for 
temperature and density are $T=300~\text{K}$
and $\rho=3.10^{28}~\text{m}^{-3}$. The thermal de Broglie 
wavelength of electrons $\lambda_{{\text{e}}}=
(\beta \hbar^2/m_{{\text{e}}})$ is 
then large compared to the mean interelectronic distance
$a=(3/(4\pi\rho))^{1/3}$, so electrons must be described by
quantum mechanics. Equivalently, thermal energy $k_BT$ is small 
compared to Fermi energy $\varepsilon_{{\text{F}}}=
\hbar^2 k_{{\text{F}}}^2/(2m_{{\text{e}}})$ with 
Fermi wavenumber $k_{{\text{F}}}=(3\pi^2\rho)^{1/3}$, 
so electrons are strongly degenerate and their typical kinetic energy 
is of order $\varepsilon_{{\text{F}}}$. Since 
$\varepsilon_{{\text{F}}}$ is small compared to the rest 
energy $m_{{\text{e}}}c^2$, relativistic effects can be omitted. 
Therefore, we have to consider the non-relativistic quantum version of 
jellium, where two electrons separated by a distance $r$ 
interact \textsl{via} the instantaneous Coulomb potential 
\be
\label{electronpot}
u_{{\text{ee}}}(r) =  \frac{e^2}{r} \; .
\ee
In addition, each electron is submitted to the 
electrostatic potential created by the ionic background, while 
the constant background self-electrostatic energy is also 
taken into account in the full interaction potential of the system.
Electrons obey to Fermi statistics since they carry an half-integer
spin. Notice that the Coulomb Hamiltonian of the present model do not depend 
on electron spins.

\subsubsection{Thomas-Fermi theory}

Similarly to the classical case described above, screening properties 
can be merely illustrated through a mean-field calculation of the equilibrium 
charge density $C_{{\text{e}}}(\br)$ surrounding an electron fixed 
at the origin, 
\be
\la{chargeOCP}
C_{{\text{e}}}(\br) = -e \(
\frac{\rho_{{\text{ee}}}^{(2)}(\mathbf{0},\br)}{\rho} 
-\rho \) \, .
\ee
In definition (\ref{chargeOCP}), $\rho_{{\text{ee}}}^{(2)}$ 
is the two-point equilibrium distribution of electrons, while 
the substracted term accounts for the ionic background density. 
The so-called Thomas-Fermi 
theory~\cite{Tho,Fer} was introduced for describing 
the electronic structure of atoms, and it can be viewed as 
the source of further density functional methods. Its application to 
the study of screening properties~\cite{ThoFer} can be rephrased 
similarly to the mean-field approach for the classical case. Namely, we 
assume that the electrons close to a given point $\br$ 
constitute an ideal gas submitted to the external 
one-body potential $-e\varphi_{{\text{e}}}(\br)$, 
where $\varphi_{{\text{e}}}(\br)$
is nothing but the electrostatic 
potential created by the electron fixed at the origin plus its polarization cloud with
charge distribution $C_{{\text{e}}}(\br)$. 
Now, according to the quantum nature of that ideal gas, its density is related 
to the external potential \textsl{via}
the Fermi-Dirac distribution instead of the Boltzmann law, \textsl{i.e.}
\be
\la{fermi}
\frac{\rho_{{\text{ee}}}^{(2)}(\mathbf{0},\br)}{\rho} =
\frac{2}{(2\pi)^3} \int \dd \bk \; \frac{1}
{\exp [\beta(\varepsilon(\bk)-e \varphi_{{\text{e}}}(\br) -\mu )] + 1} \; .
\ee 
In Fermi-Dirac expression (\ref{fermi}), $\varepsilon(\bk)=\hbar^2 k^2/(2m_{{\text{e}}})$ 
is the non-relativistic kinetic energy of a plane wave 
with wavenumber $\bk$, while the 
chemical potential $\mu$, taken constant across the system, 
is naturally determined by the bulk condition
\be
\la{fermibulk}
\rho = \frac{2}{(2\pi)^3} \int \dd \bk \; \frac{1}
{\exp [\beta(\varepsilon(\bk) -\mu )] + 1} \; .
\ee 
Assuming that polarization effects vanish at large 
distances, we set $\rho_{{\text{ee}}}^{(2)}(\mathbf{0},\br) 
\rightarrow  \rho^2$ when $r \rightarrow \infty$.
The corresponding limit form of equation 
(\ref{fermi}) then implies the boundary condition 
$\varphi_{{\text{e}}}(\br) \rightarrow 0$. 

\bigskip

As argued for the Boltzmann factor involved in Debye theory, 
the linearization of Fermi-Dirac distribution (\ref{fermi}) 
with respect to $\varphi_{{\text{e}}}(\br)$, is consistent 
with the present mean-field treatment, which also 
requires that Coulomb interactions are small 
perturbations. If we set $\varphi_{{\text{e}}}(\br)=-e\phi_{{\text{TF}}}(\br)$,
we then obtain an Hemholtz equation for $\phi_{{\text{TF}}}(\br)$ which 
is identical to its classical counterpart (\ref{helmoltz}) for 
$\phi_{{\text{D}}}(\br)$, except for the replacements 
$\epsilon_{{\text{w}}} \rightarrow 1$ and 
$\kappa_{{\text{D}}} \rightarrow \kappa_{{\text{TF}}}$
where the Thomas-Fermi wavenumber $\kappa_{{\text{TF}}}$ is defined by
\be
\la{kappaTF}
\kappa_{{\text{TF}}}^2= 4 \pi e^2 
\frac{\partial \rho }{\partial \mu} (\beta,\mu) \; .
\ee
The resulting mean-field expression for charge 
distribution $C_{{\text{e}}}(\br)$ reads
\be
\la{TF}
-e \kappa_{{\text{TF}}}^2 \frac 
{\exp (-\kappa_{{\text{TF}}} r)}{4\pi r} \; .
\ee
Thus, the mean-field approach again
predicts an exponential screening. Now the screening 
length $\lambda_{{\text{TF}}}$ takes into account 
degeneracy effects controlled by dimensionless 
parameter $\varepsilon_{{\text{F}}}/k_BT$. In the limit of weak degeneracy, which 
can be obtained by setting $\mu \rightarrow -\infty$ at fixed $\beta$, 
the density vanishes as $\rho \sim 2 \exp(\beta \mu)/(2\pi\lambda_{{\text{e}}})^{3/2}$, so
$\lambda_{{\text{TF}}}$ does reduce to its classical 
Debye form $(4\pi\beta e^2 \rho)^{-1/2}$. In the opposite strong-degeneracy limit, 
obtained by setting $\mu \rightarrow +\infty$ at fixed $\beta$, the density 
diverges as $\rho \sim (2m_{{\text{e}}} \mu )^{3/2}/(3\pi^2)$ and 
we find 
\be
\la{TFlength}
\lambda_{{\text{TF}}} \sim \( \frac{\pi^2}{144} \)^{1/6} 
\( \frac{\hbar^2  a}{m_{{\text{e}}}e^2} \)^{1/2} \; .
\ee

\subsubsection{Expected validity regime}

As quoted above, the validity of Thomas-Fermi theory requires 
weak-coupling conditions. Of particular interest is
the regime of strong degeneracy, where 
the relevant kinetic energy is  the Fermi 
energy $\varepsilon_{{\text{F}}}$. The corresponding weak-coupling condition 
reads $e^2/(a\varepsilon_{{\text{F}}}) \ll 1$. That strongly-degenerate 
weakly-coupled regime is reached at sufficiently high densities for a given 
temperature. Indeed, when $\rho \rightarrow +\infty$ at fixed $\beta$, 
Fermi energy obviously becomes larger than classical thermal energy, 
while Coulomb energy of order $\rho^{1/3}$ grows slower than kinetic
Fermi energy of order $\rho^{2/3}$. In that high-density limit, notice 
that $\lambda_{{\text{TF}}}$, given by expression (\ref{TFlength}), 
becomes large compared to $a$, so the number of electrons inside 
the screening sphere with radius $\lambda_{{\text{TF}}}$ is 
indeed large.

\bigskip

In order to derive corrections to the mean-field approach, 
it is necessary to use the formalism of many-body 
perturbation theory~\cite{FetWal}. In that framework, 
the mean-field ideas are recasted into the celebrated 
Random Phase Approximation (RPA), which 
is close to Thomas-Fermi theory. In particular, RPA
also predicts an exponential decay 
of charge correlations. Systematic corrections to RPA can be 
derived by taking into account suitable Feynman graphs. 
This leads to asymptotic high-density expansions of 
thermodynamical quantities, like the pressure or the 
internal energy, beyond their ideal strongly-degenerate forms. 

\bigskip

For the strongly-degenerate electron gas described above, 
the Coulomb energy is not small compared to the Fermi 
energy. Therefore the system is not weakly coupled, and the application 
of Thomas-Fermi theory is rather questionable. In particular, 
we find that $\lambda_{{\text{TF}}} \simeq 0.1~\text{nm}$ is 
smaller than $a \simeq 0.2~\text{nm}$, in violation of 
the mean-field validity condition $\lambda_{{\text{TF}}} \gg a$ 
\footnote{
Also, in the present calculation, we omit all effects present in 
real condensed matter, like those arising from either
the band structure of the electronic spectrum, or the 
electron-phonon coupling~\cite{ThoFer,SolSta}. Despite its obvious drawbacks, 
Thomas-Fermi calculation suggests that the electron-electron 
effective potential is shorter ranged than the bare Coulomb interaction.}.

\subsubsection{Doubts about mean-field predictions}

If mean-field approaches like RPA may provide the 
leading behaviour of integrated quantities involving 
$C_{{\text{e}}}(\br)$ in the high-density limit 
$\rho \rightarrow +\infty$ at fixed $\beta$, their reliability 
for $C_{{\text{e}}}(\br)$ itself is not uniform with respect 
to distance $r$. At short distances, the divergency of Thomas-Fermi
expression (\ref{TF}) for $C_{{\text{e}}}(\br)$ is 
obviously unphysical, like for its classical counterpart 
in the framework of Debye theory. This is due first 
to the strong repulsion embedded in $\varphi_{{\text{e}}}(\br)$
when $r \rightarrow 0$ which cannot be treated perturbatively.  
Also TF theory does not account for the effective 
repulsion induced by Pauli principle when 
both electrons located at $\mathbf{0}$ and $\br$ have the same 
spin orientations. At large distances, 
and contrarily to the classical case, mean-field 
predictions are also doubtful, as argued below. 

\bigskip

In writing the Fermi-Dirac distribution (\ref{fermi}), 
a very important implicit assumption is made. Indeed, 
potential $\varphi_{{\text{e}}}(\br)$ is treated as 
a constant, so the energy levels 
are merely shifted by that constant from their 
corresponding purely kinetic expression $\varepsilon(\bk)=\hbar^2 k^2/(2m_{{\text{e}}})$. 
This amounts to neglect the spatial variations of $\varphi_{{\text{e}}}(\br)$, 
or in other words to omit diffraction effects 
arising from the non-commutativity of the kinetic and potential parts of 
Hamiltonian $-\hbar^2 \Delta/(2m_{{\text{e}}}) + \varphi_{{\text{e}}}(\br)$.  
Then, we are left with a purely configurational static problem, like 
in the classical case, and not surprisingly this leads also to an exponential 
screening. However, that result is rather questionable since an 
intrinsic specificity of quantum mechanics has been erased from the 
start!  

\bigskip

The occurence of quantum diffraction in equilibrium statistical weights 
is a signature of dynamical effects, which are consequently still 
present at equilibrium. Now, and contrarily to the 
classical case, dynamical fluctuations combined to inertia effects
should prevent instantaneous perfect arrangments of equilibrium screening clouds.
Thus, the presence of partially screened multipolar interactions 
can be anticipated, and this might lead to a breakdown 
of exponential screening in equilibrium 
correlations. That quite plausible scenario 
has been first conjectured in Ref.~\cite{BrySei}, according to
the presence of algebraic tails in some imaginary-time 
Green functions. Notice that an analogous scenario occurs 
for classical time-displaced correlations~\cite{AlaMar2}.

\subsection{Recombination and van der Waals interactions} \label{ssrecombination}

\subsubsection{Hydrogen in the Sun}

The sun is mainly made of hydrogen and helium, while heavier elements 
like oxygen, carbon, iron,...are less abundant and contribute to a 
small fraction of its total mass. Here, we will focus our study 
on hydrogen, and we will discard its interactions with all the other 
elements. Also, we do not consider complicated phenomena at work inside 
the Sun, like radiation transfer or convection for instance. For 
our purpose, it is sufficient to assume that some local thermodynamical equilibrium 
is reached at any place, and then the central question is 
to determine the corresponding chemical composition of an 
hydrogen plasma in terms of ionized and recombined entities.

\bigskip

\begin{figure}
\psfragscanon
\psfrag{a}[c][c]{\large{\textsl{Core}}}
\psfrag{b}[c][c]{\large{\textsl{Intermediate layers}}}
\psfrag{c}[c][c]{\large{\textsl{Photosphere}}}
\psfrag{S}[c][c]{\Large{\textbf{Sun}}}
\includegraphics[width=0.6\textwidth]{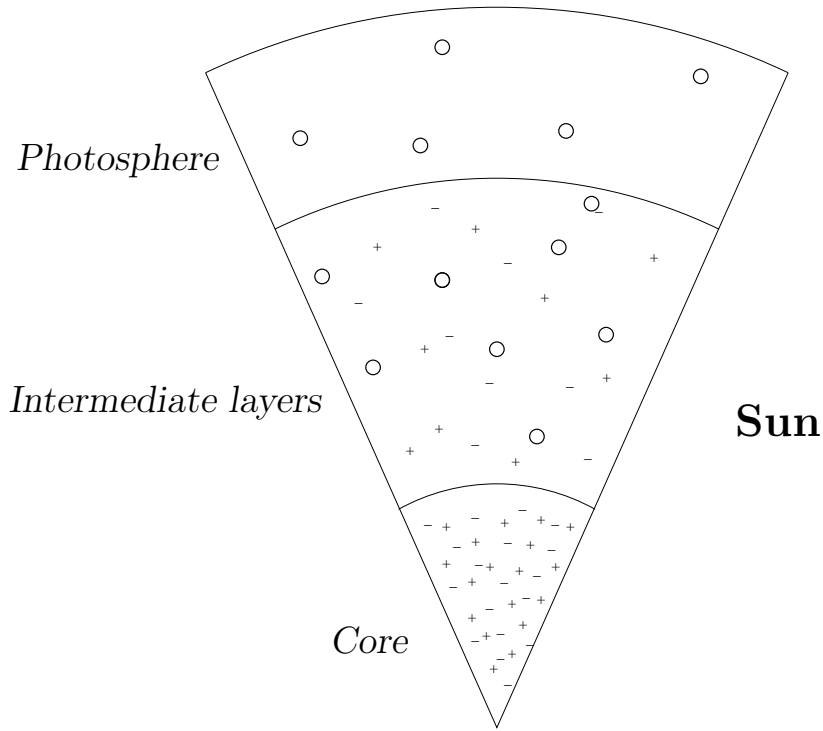}
\caption{\label{Sun}A schematic view of the Sun. Signs $+$ 
and $-$ represent ionized protons and ionized electrons 
respectively, while circles represent hydrogen atoms.}
\end{figure}

In the core of the Sun, temperature $T$ is of the order of 
$10^6~\text{K}$ and density $\rho$ is of order 
$1~\text{g/cc}$. Since $k_BT \simeq 90~\text{eV}$ is much larger than 
the Rydberg energy of order $13.6~\text{eV}$, atoms, molecules, as 
well as other recombined entities are then fully ionized. The 
hydrogen gas is then mainly composed of free protons and of 
free electrons. In intermediate layers, temperature and density 
decrease as the distance to the core increases. This favors 
recombination of protons and electrons into atoms, 
and hydrogen then behaves as a partially ionized atomic gas. 
In the photosphere, $T$ is of order 
$6.10^3~\text{K}$ while $\rho$ is of order $10^{-7}~\text{g/cc}$. 
Under such cool ($k_BT \simeq 0.5~\text{eV}$ much 
smaller than Rydberg energy) and diluted  
(the mass density of Earth atmosphere is of order 
$10^{-3}~\text{g/cc}$) conditions, the full proton-electron 
recombination is achieved, and hydrogen reduces to an atomic gas.
That recombination process is schematically represented in Fig.~\ref{Sun}.

\bigskip

A proper description of above recombination process within 
statistical mechanics requires to account for both 
quantum mechanics and Coulomb interactions as detailed 
in the next Section. Here, we list several natural 
problems, and we present simple preliminary arguments 
which are particularly useful for a better 
understanding of more sophisticated analysis 
carried out in the framework of the many-body problem.

\subsubsection{How to define an hydrogen atom?}

In quantum mechanics, all particle positions 
are entangled in a global wavefunction describing the whole system.
If we consider an eigenfunction of the total Hamiltonian, 
no two-body wavefunction describing an hydrogen atom can be factored out. 
In that context, an unambiguous definition of such an atom 
is only possible in the double limit $\rho \to 0$ and 
$T \to 0 $. Indeed, the zero-density limit ensures 
that once a proton has married an electron, all other 
neighbours are very far apart. Also, the zero-temperature 
limit guarantees that the atom stays in its groudstate and 
no thermal ionization occurs. Under such infinitely 
diluted and cold conditions, it can be reasonably 
expected that a single hydrogen atom in the vacuum 
naturally emerges.

\subsubsection{How protons and electrons recombine into atoms?}

The formation of atoms in a gas of protons and electrons has 
been studied long ago by Saha~\cite{Saha}, in the framework of 
the chemical picture. He considered protons, electrons and atoms 
as three independent species with their own chemical potentials. 
If all interactions are neglected and if each species is viewed as 
a classical ideal gas, the mass action law associated with the recombination 
equation $\text{p} + \text{e} \rightleftharpoons \text{H}$, reads
in terms of species ideal densities 
\be
\la{Saha}
\frac{\rho_{{\text{p}}}^{{(\text{id})}}
\rho_{{\text{e}}}^{{(\text{id})}}}
{\rho_{{\text{at}}}^{{(\text{id})}}} =
\frac{\exp(\beta E_{{\text{H}}})}{(2\pi\lambda_{{\text{pe}}}^2)^{3/2}} \; ,
\ee
where $E_{{\text{H}}}$ is the 
groundstate energy of an hydrogen atom, 
$E_{{\text{H}}}=-me^4/(2\hbar^2)$, with 
reduced mass $m=(m_{{\text{p}}}m_{{\text{e}}})/(m_{{\text{p}}}+m_{{\text{e}}})$
and associated de Broglie wavelength $\lambda_{{\text{pe}}}=(\beta\hbar^2/m)^{1/2}$.
The temperature-dependent reaction constant in the r.h.s. of 
mass action law (\ref{Saha}), 
goes to zero when $T$ vanishes, while it explodes when $T$ diverges. 
Therefore, and as expected, low temperatures favor atomic recombination, 
while high temperatures favor ionization.

\bigskip

At given temperature and density, the chemical composition of 
the hydrogen gas is determined by adding to mass action law (\ref{Saha})
the neutrality rule, $\rho_{{\text{p}}}^{{(\text{id})}}=
\rho_{{\text{e}}}^{{(\text{id})}}$, as well as the total 
density constraint, $\rho_{{\text{p}}}^{{(\text{id})}} +
\rho_{{\text{at}}}^{{(\text{id})}}=\rho$. A straightforward 
calculation then provides 
\be
\la{Sahadensity1}
\rho_{{\text{p}}}^{{(\text{id})}}=
\rho_{{\text{e}}}^{{(\text{id})}}=
\rho^{\ast}\(\sqrt{1+2\rho/\rho^{\ast}}-1\)
\ee
and 
\be
\la{Sahadensity2}
\rho_{{\text{at}}}^{{(\text{id})}}=
\rho^{\ast}\(1+ \rho/\rho^{\ast} -\sqrt{1+2\rho/\rho^{\ast}} \) \; 
\ee
with temperature-dependent density 
\be
\la{rhostar}
\rho^{\ast}=\frac{\exp(\beta E_{{\text{H}}})}
{2(2\pi\lambda_{{\text{pe}}}^2)^{3/2}} \; .
\ee
According to those simple formulae, hydrogen becomes fully ionized 
for $\rho \ll \rho^{\ast}$, while it reduces to an atomic gas for 
$\rho \gg \rho^{\ast}$. This explains the behaviour of hydrogen inside 
the Sun, since the corresponding ratio $\rho/\rho^{\ast}$ varies 
from $10^{-3}$ in the core to $10^{7}$ in the photosphere.

\subsubsection{Why atoms and not molecules?}

\begin{figure}
\psfragscanon
\psfrag{Sp}[c][c]{{\textsl{Sun photosphere}}}
\psfrag{Sc}[c][c]{{\textsl{Sun core}}}
\psfrag{Ea}[c][c]{{\textsl{Earth atmosphere}}}
\psfrag{rhostar}[c][c]{$\rho^{\ast}(\beta)$}
\psfrag{0}[c][c]{$0$}
\psfrag{a}[c][c]{$\beta$}
\psfrag{b}[c][c]{$\log (\rho a_{\text{B}}^3)$}
\psfrag{P}[c][c]{\large{\textbf{Plasma}}}
\psfrag{A}[c][c]{\large{\textbf{Atomic gas}}}
\psfrag{M}[c][c]{\large{\textbf{Molecular gas}}}
\includegraphics[width=0.6\textwidth]{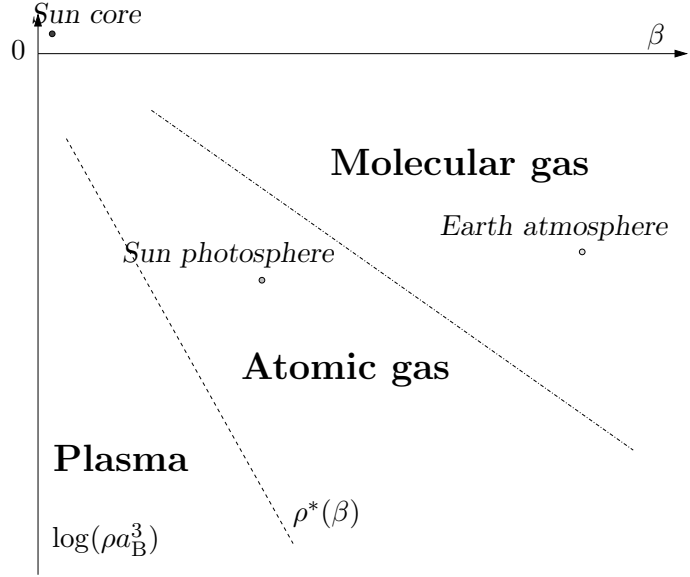}
\caption{\label{Phase}Phase diagram of hydrogen at low densities. 
Regions where either ionized charges, atoms or molecules prevail, 
are separated by dashed ($\rho^{\ast}(\beta)$) and dash-dotted lines, in the 
vicinity of which the respective dissociation/recombination processes 
take place.}
\end{figure}

In Earth atmosphere, hydrogen is a molecular gas and no atoms are 
present. This is oftenly explained by invoking that 
molecule $\text{H}_2$ is more stable than atom $\text{H}$, namely 
$E_{{\text{H}_2}} < 2 \; E_{{\text{H}}}$, where 
$E_{{\text{H}_2}} $ is the molecule groundstate energy. However, we stress 
that such an argument strictly applies at zero temperature, 
where only energy $U^{\ast}$ intervene. At 
finite temperatures, $T > 0$, entropy $S^{\ast}$ must be also 
considered in order to determine the chemical composition 
of the system. For a given number $N=N_p=N_e$ of protons and electrons 
enclosed in a box with volume $\Lambda$ at a given temperature $T$, 
the entropy of the atomic gas is larger than that of 
its molecular counterpart, because the atom number 
$N_{{\text{H}}}=N/2$ is twice the molecule number 
$N_{{\text{H}_2}}=N/4$. Thus, in the minimization of 
the thermodynamic potential $\( U^{\ast}-TS^{\ast} \)$ at fixed $T$, $\Lambda$ and $N$, 
entropy contribution favors the formation of atoms~\footnote{
The fractions of atoms and molecules can be determined 
within a simplified mixture model of non-interacting classical particles 
with an internal energy $E_{{\text{H}_2}}$ or $E_{{\text{H}}}$. 
The corresponding minimization equation then reduces to 
the usual mass action law associated with dissociation
reaction $\text{H}_2 \rightleftharpoons 2\text{H}$.}. That effect becomes 
more important as $T$ increases or $\rho=N/\Lambda$ decreases. In 
Earth atmosphere, temperature is rather low and density is rather high so 
only molecules are formed. In the Sun, temperature is sufficiently large and 
density sufficiently low so atoms prevail. The phase diagram of 
hydrogen at low densities is schematically drawn in Fig.~\ref{Phase}.

\subsubsection{What about the internal atomic partition function?}

At finite temperatures, not only the atomic groundstate contributes 
to thermodynamic quantities, but also excited states.
When $T$ vanishes, an atom can be considered as frozen in 
its groundstate because of the finite gap between 
$E_{{\text{H}}}$ and the first excited level $E_{{\text{H}}}/4$. 
The corresponding contribution to equilibrium quantities
merely involves Boltzmann factor $\exp(-\beta E_{{\text{H}}})$. 
At finite temperatures, contributions of excited states
must also be taken into account. For that purpose, it is natural 
to introduce the internal atomic partition function in the vacuum
\be
\la{internalpartition}
Z_{{\text{H}}}^{{(\text{vac})}} = 4 \sum_{n=1}^{\infty} \; n^2 \;
\exp(-\beta E_{{\text{H}}}/n^2) \; .
\ee
In that definition, factor $4$ arises from proton and electron spins 
degeneracy, factor $n^2$ accounts for orbital degeneracy 
of the level $n$ with energy $E_{{\text{H}}}/n^2$, and 
summation extends over all boundstates up to $n=\infty$.

\bigskip

Obviously $Z_{{\text{H}}}^{{(\text{vac})}}$ diverges because of the contributions 
of Rydberg states with $n \to \infty$. In the framework of the many-body 
problem, that divergency is usually expected to be cured 
thanks to screening effects, according to the following rough argument. 
At finite $T$ and $\rho$, there is a finite amount of free charges, 
so the proton-electron potential inside an hydrogen atom 
decays faster than $1/r$ at distances $r$ larger than screening length 
$\lambda_{{\text{S}}}$. Thus, there is no accumulation of 
atomic boundstates near the continuum edge at $E=0$, and 
all Rydberg states with $n$ sufficiently large are suppressed. 
Since the spatial extension of such states is of order $n^2 a_{{\text{B}}}$
with Bohr radius $a_{{\text{B}}}=\hbar^2/(me^2)$, 
the infinite sum (\ref{internalpartition})
has to be truncated up to 
$p_{{\text{S}}} \sim (\lambda_{{\text{S}}}/a_{{\text{B}}})^{1/2}$. This 
provides a non-divergent atomic partition function,
\be
\la{internalpartitionbis}
Z_{{\text{H}}}^{{(\text{scr})}} = 4 \sum_{n=1}^{p_{{\text{S}}}} \; n^2 \;
\exp(-\beta E_{{\text{H}}}/n^2) \; ,
\ee
which accounts for screening effects.

\subsubsection{Are van der Waals interactions screened?}

As guessed by van der Waals in his thesis, atoms or molecules 
attract themselves at large distances. For atoms with no 
permanent dipole, that interaction is due to quantum fluctuations 
as first shown by London~\cite{Lon}. Let us sketch the corresponding calculation 
for two hydrogen atoms in their groundstate. Each atom $i$ 
($i=1,2$) is described by an Hamiltonian $H_{{\text{at}}}^{(i)}$, 
and the full Hamiltonian of both atoms reads
\be
\la{hamiltonianatoms}
H=H_{{\text{at}}}^{(1)}+ V_{{\text{at}},{\text{at}}} +H_{{\text{at}}}^{(2)}
\ee
where $V_{{\text{at}},{\text{at}}}$ is 
the electrostatic interaction between the two atoms. For large 
separations $R$ between such atoms, namely $R \gg a_{{\text{B}}}$, 
that potential becomes dipolar because of atom charge neutrality, 
\textsl{i.e.}
\be
\la{atomatom}
V_{{\text{at}},{\text{at}}} 
\sim \( \bp_1 \cdot \na \)\( \bp_2 \cdot \na \)\frac{1}{R} 
\;\;\;\text{when}\;\;\; R \to \infty \; ,
\ee
where $\bp_i$ is the instantaneous electrical dipole carried by 
atom $i$. At such distances, when 
determining the groundstate energy of the whole two-atoms system,
$V_{{\text{at}},{\text{at}}}$ 
can be treated as a small perturbation with respect to 
$H_{{\text{at}}}^{(1)}+ H_{{\text{at}}}^{(2)}$. The 
first-order correction vanishes because the quantum average of 
$\bp_i$ in the unperturbed groundstate is zero. The second-order 
correction can be reinterpreted as an atom-atom effective 
potential which reads~\cite{Sch}
\be
\la{vdWpotential}
U_{{\text{vdW}}}(R) = -\frac{C_{{\text{vdW}}}}{|E_{{\text{H}}}|} \;
\frac{e^4 a_{{\text{B}}}^4}{R^6} \; .
\ee
In that expression, $C_{{\text{vdW}}}$ is a pure numerical constant 
which is positive, so van der Waals potential $U_{{\text{vdW}}}(R)$ is 
attractive. That potential appears to be generated by quantum 
fluctuations of the instantaneous $1/R^3$-dipolar interaction 
between two atoms : this explains the $1/R^6$-decay of 
$U_{{\text{vdW}}}(R)$ at large distances, which merely 
follows by taking the square of $1/R^3$. Eventually, we check 
\textsl{a posteriori} that use of perturbation theory for 
deriving formula (\ref{vdWpotential}) is indeed justified
if $R \gg a_{{\text{B}}}$, since 
$|U_{{\text{vdW}}}(R)|/|E_H|$ is proportional to $(a_{{\text{B}}}/R)^6$.

\bigskip

Above derivation is performed for two isolated atoms 
at zero temperature in the vacuum. 
In the framework of the many-body system at finite 
$T$ and $\rho$, besides the above difficulties in 
defining properly atoms, 
an interesting question concerns the
effects of free charges on van der Waals interactions. 
Since the instantaneous atom-atom potential $V_{{\text{at}},{\text{at}}}$ 
is of purely Coulombic origin, screening mechanisms should 
modify $U_{{\text{vdW}}}(R)$ at distances $R$ larger than 
screening length $\lambda_{{\text{S}}}$. 
A very crude and naive application 
of the classical recipe would lead to an exponential decay of 
the effective atom-atom interactions. 
Nevertheless, and as it has been anticipated in the 
construction of the phenomenological Derjaguin-Landau-Verwey-Overbeek 
theory for mixtures of charged and 
neutral fluids~\cite{DL,VO,DLVO}, screening of 
van der Waals interactions should be less 
efficient than that prediction because of their underlying dynamical character. 
Hence, it is argued that the fast electronic motion 
embedded in quantum fluctuations of atom dipoles cannot 
be instantaneously followed by free charges because 
of inertia. According to that heuristic argument, 
it is oftenly assumed that $U_{{\text{vdW}}}(R)$ 
is not screened at all. Notice
that such argument is close to the one 
presented previously which suggests a breakdown of 
exponential screening in the quantum case.

\section{Some exact results within the screened cluster representation}

\subsection{Fundamental issues and the hydrogen quantum plasma} \label{ssissues}

In the previous introductory part of that lecture,
various fundamental issues have been raised, namely

\bigskip

\noindent $\bullet$ \textsl{Nature of screening in quantum systems}

\bigskip

\noindent $\bullet$ \textsl{Introduction of recombined entities in the many-body problem}

\bigskip

\noindent $\bullet$ \textsl{Modification of van der Waals interactions due to free charges}

\bigskip

\noindent According to the preliminary analysis and discussions, 
it clearly appeared that a proper and reliable account of 
the relevant mechanisms requires the introduction of an elementary description of 
matter in terms of a quantum Coulomb system made with point 
nuclei and point electrons. For the sake of simplicity, and also because 
of its wide interest for both conceptual purposes and practical applications, 
here we will restrict ourselves to the hydrogen plasma viewed as a system 
of quantum point particles which are either protons or electrons, 
interacting via the instantaneous Coulomb potential. 
Protons and electrons have respective charges, masses, and spins, 
$e_{{\text{p}}}=e$ and $e_{{\text{e}}}=-e$, 
$m_{{\text{p}}}$ and $m_{{\text{e}}}$, 
$\sigma_{{\text{p}}}=\sigma_{{\text{e}}}=1/2$. 
In the present non-relativistic limit, 
the corresponding Hamiltonian for $N=N_{{\text{p}}}+N_{{\text{e}}}$ particles reads
\begin{equation}
H_{N_{{\text{p}}},N_{{\text{e}}}} = -\sum_{i=1}^N {\hbar^2 \over 2m_{\alpha_i}} \Delta_i + 
{1 \over 2}\sum_{i \neq j} e_{\alpha_i} e_{\alpha_j} v(|\textbf{x}_i - \textbf{x}_j|)
\label{deuxun}
\end{equation}
with $v(r)=1/r$, $(\alpha_i=\text{p},\text{e})$ the species of the $i$th particle,
and $\Delta_i$ the Laplacian with respect to its position $\textbf{x}_i$. 
In the following, we present several exact results for the 
equilibrium properties of that system, in relation with above 
fundamental issues.

\bigskip

First, in Section~\ref{sshydrogen}, we 
recall various rigorous results about the 
existence of the thermodynamic limit, as well as about
the partially ionized atomic regime. Then, in Section~\ref{sspath}, we 
introduce the path integral representation, in the simple case of a single particle submitted 
to an external potential for the sake of pedagogy. Application of that tool 
to the many-body quantum system leads to the 
introduction of an equivalent classical system made with extended 
objects called loops, which interact \textsl{via} a two-body potential
equal to some average along their shapes of the genuine two-body 
particle interaction. The equilibrium quantities of the 
quantum plasma can then be represented by Mayer-like 
diagrammatical series in the loops world. In Section~\ref{ssSCR}, 
we describe the resummation machinery which exactly
transforms the sum of divergent Mayer graphs into 
a sum of convergent graphs with a similar structure, 
the so-called screened cluster representation (SCR). 
Long-range Coulomb divergences are 
removed \textsl{via} systematic 
chain resummations, which amount to introduce 
a screened effective potential between loops. Further reorganizations
lead to the introduction of particle clusters with finite 
statistical weights, which include the contributions of
familiar chemical species. Exact asymptotic expansions 
in the partially ionized atomic regime can be derived within 
that SCR, for both the equation of state and particle correlations, 
as decribed in Section~\ref{applications}. Eventually, 
we summarize in Section~\ref{conclusion} the main answers to the 
fundamental issues quoted above, which can be inferred from 
the present analysis of the hydrogen plasma.

\subsection{Rigorous results} \label{sshydrogen}

\subsubsection{Thermodynamic limit}

Let us consider the hydrogen plasma in the framework of the 
grand-canonical ensemble. The system is enclosed in a box with volume $\Lambda$, 
in contact with a thermostat at temperature $T$ and a reservoir of particles 
that fixes the chemical potentials equal to $\mu_{{\text{p}}}$ and $\mu_{{\text{e}}}$ 
for protons and electrons respectively. The corresponding 
grand-partition function reads 
\be
\la{gpf}
\Xi_{\Lambda}=\text{Tr}_{\Lambda}\exp \(-\beta (H_{N_{{\text{p}}},N_{{\text{e}}}} -
\mu_{{\text{p}}}N_{{\text{p}}}-\mu_{{\text{e}}}N_{{\text{e}}})   \) \; .
\ee 
The trace $\text{Tr}_{\Lambda}$ is taken over a complete basis of  
N-body wavefunctions, which are symmetrized according 
to Fermi statistics and satisfy Dirichlet boundary conditions at 
the surface of the box, while particle numbers 
$N_{{\text{p}}}$ and $N_{{\text{e}}}$ 
vary from $0$ to $\infty$~\footnote{That standard definition of 
the grand-canonical ensemble may be viewed as rather artificial in the sense that 
Dirichlet boundary conditions do not describe the physical situation 
which is considered in the Gibbs construction, 
where particles can freely go in and out the volume $\Lambda$ 
immersed inside a much larger container. Nevertheless, in the 
thermodynamic limit $\Lambda \to \infty$, boundary conditions 
become irrelevant for bulk quantities. Therefore, for our purpose it is legitimate 
to use definition (\ref{gpf}), where particle numbers 
are varied by some hidden operator.}. 

\bigskip

The thermodynamic limit (TL) is defined 
by the limit $\Lambda \to \infty$ at fixed temperature 
and fixed chemical potentials. In that limit, Lieb and Lebowitz~\cite{LieLeb} 
proved the existence of a well-defined bulk equilibrium state with the 
right extensive properties expected from macroscopic thermodynamics. 
That state is overall neutral, namely the average particle densities 
$\rho_{{\text{p}},\Lambda}=\langle N_{{\text{p}}}\rangle/\Lambda$ and 
$\rho_{{\text{e}},\Lambda}=\langle N_{{\text{e}}}\rangle/\Lambda$ 
become identical in the TL, 
\be
\la{neutralitygce} 
\lim_{{\text{TL}}}\rho_{{\text{p}},\Lambda}=
\lim_{{\text{TL}}}\rho_{{\text{e}},\Lambda}= \rho \; .
\ee
That common particle density depends only on temperature $T$ 
and on the mean chemical potential
\begin{equation}
\mu=\frac{\mu_{{\text{p}}} + \mu_{{\text{e}}}}{2} \; , 
\label{meanmu}
\end{equation}
while the difference $\nu=(\mu_{{\text{e}}}-\mu_{{\text{p}}})/2$ is not relevant.
Pressure $P$ defined by the standard formula
\be
\la{pressure}
P= k_BT \; \lim_{{\text{TL}}} \frac{ \ln \Xi_{\Lambda}}{\Lambda} 
\ee
is also a function of the two parameters $T$ and $\mu$, as well 
as all the other intensive thermodynamical quantities. 
Moreover, bulk properties no longer depend on the choosen boundary conditions, 
and all statistical ensembles become equivalent in the TL. In particular, 
grand-canonical expression (\ref{pressure}) for pressure $P(T,\mu)$
does coincide with its canonical counterpart $P(T,\rho)$ evaluated at
density $\rho=\rho(T,\mu)$.

\bigskip

Lieb and Lebowitz proof involves sophisticated mathematical analysis which 
will not be detailed here, of course! However, it is instructive to 
present the main ideas underlying the derivations. In fact, the central 
difficulties are related to the singular behaviours of the 
Coulomb potential at both short and large distances, as already quoted 
in Section~\ref{ssexamples}. The short-distance collapse is avoided 
thanks to a combination of the Heisenberg uncertainty principle 
and of Fermi statistics. According to that principle, the $1/r$-singularity 
is smeared out, so the groundstate energy of $N$ particles is finite, instead 
of $-\infty$ when opposite charges are present. Fermi statistics 
prevent the modulus of negative groundstate energies to grow faster than $N$. 
At a mathematical level, this is rephrased in the celebrated 
H-stability theorem~\cite{DysLen},
\be
\la{stability}
H_{N_{{\text{p}}},N_{{\text{e}}}} > 
-B\; (N_{{\text{p}}}+N_{{\text{e}}}) 
\ee
where $B$ is a strictly positive constant. The 
large-distance explosion is prevented by screening mechanisms, 
the mathematical formulation of which is embedded 
in the cheese theorem. That theorem relies on a suitable 
partition of space in neutral spheres surrounding the charges, 
which applies for most probable configurations. Indeed, such
configurations are overall neutral in the bulk while  
excess charges are repelled on the boundaries,
as suggested by macroscopic electrostatics.   
The harmonicity of Coulomb potential then plays a crucial role, since 
according to Gauss theorem, two non-overlapping neutral spheres 
do not interact. This implies that dangerous long-range contributions 
do not intervene in the TL. Both cheese and H-stability theorems are the key ingredients 
of the proof.

\subsubsection{Partially ionized atomic regime}

In Section~\ref{ssrecombination}, we showed how, according to phenomenological 
Saha theory, a finite fraction of protons and electrons 
might recombine into hydrogen atoms, forming a partially ionized atomic gas. 
Also, we argued that such a regime should be attained when 
both temperature and density are sufficiently low 
so that individual atoms in their groundstate can form, namely $kT \ll |E_H|$ 
and $a \gg a_B$ with $a=(3/(4\pi\rho))^{1/3}$ the mean inter-particle distance. 

\bigskip

The above Saha prediction has been rigorously proved 
within a suitable scaling limit where both temperature and 
density vanish in a related way. That scaling limit is introduced in the framework 
of the grand-canonical ensemble : temperature $T$ goes to zero while 
average chemical potential $\mu$ approaches the value $E_H$ 
with a definite slope, according to parametrization~\cite{MacMar}
\begin{equation}
\label{mu_gamma}
\mu =  E_{{\text{H}}} + k_B T [\ln(\gamma) + \ln((m/M)^{3/4}/4) ]
\end{equation}
with fixed dimensionless parameter $\gamma$ and 
$M=m_{{\text{p}}}+m_{{\text{e}}}$. Then density
$\rho$ indeed vanishes with $T$, and it behaves as
\be
\la{rhoSL}
\rho = \rho^{\ast}\gamma (1+\frac{\gamma}{2}) \( 1+{\cal O}(e^{-c\beta}) \) \; ,
\ee
with $c$ a postive constant and temperature-dependent 
density $\rho^{\ast}$ given by formula (\ref{rhostar}). The 
corresponding asymptotic behaviour of pressure $P$ reads 
\be
\la{pressureSL}
\beta P = \rho^{\ast}\gamma (2+\frac{\gamma}{2}) \( 1+{\cal O}(e^{-c\beta}) \) \; .
\end{equation}
Thus, if we discard the exponentially small corrections embedded in 
the ${\cal O}(e^{-c\beta})$-terms, the corresponding leading behaviours of 
density and pressure can be reinterpreted in terms of ideal densities of 
protons, electrons and atoms, identified to 
\be
\la{rhoideal1}
\rho_{{\text{p}}}^{{(\text{id})}}=
\rho_{{\text{e}}}^{{(\text{id})}}=
\rho^{\ast} \gamma
\ee
and 
\be
\la{rhoideal2}
\rho_{{\text{at}}}^{{(\text{id})}}=
\rho^{\ast}\frac{\gamma^2}{2} \; .
\ee
Such expressions do coincide with their Saha counterparts
(\ref{Sahadensity1}) and (\ref{Sahadensity2}), as it can easily checked 
by eliminating $\gamma$ in favor of $\rho$. Also, the pressure indeed 
reduces to that of an ideal mixture of protons, electrons and atoms, and  
it eventually reads
\be
\la{Sahapressure}
\beta P_{{\text{Saha}}} = \rho^{\ast}\(\rho/\rho^{\ast} 
+\sqrt{1+2\rho/\rho^{\ast}}-1 \) \; .
\ee
That rigorous derivation constitute another \textsl{tour de force} in 
mathematical physics, which has been implemented 
through successive works~\cite{Con,MacMar}
after the pionneering paper about the purely atomic limit 
by Fefferman~\cite{Fef}. Below, we sketch various estimations and 
arguments which provide a simple physical 
understanding of the result.

\bigskip

First, the above scaling limit 
defines the proper energy-entropy balance which 
ensures the emergence of the contributions of ionized protons, ionized electrons and 
hydrogen atoms in grand-partition function (\ref{gpf}). That balance results from the 
competition between the fugacity factors $\exp (\beta \mu)$ which decay exponentially fast as 
$\exp(\beta E_{{\text{H}}})$, and the groundstate Boltzmann factors 
which may explode exponentially fast. The 
relevant ideal contributions of ionized protons and ionized electrons are 
merely given by terms $(N_{{\text{p}}}=1,N_{{\text{e}}}=0)$ and
$(N_{{\text{p}}}=0,N_{{\text{e}}}=1)$, and they are of order 
$\exp(\beta E_{{\text{H}}})$. The corresponding atomic contribution,
easily picked out in term $(N_{{\text{p}}}=1,N_{{\text{e}}}=1)$, 
is also of order $\exp(\beta E_{{\text{H}}})$ since it involves 
the product of entropy factor $\exp (2\beta \mu)$ times the atomic
Boltzmann factor $\exp(-\beta E_{{\text{H}}})$. Thanks to  
remarkable inequalities~\footnote{In fact, those inequalities have been conjectured 
by Fefferman~\cite{Fef}, but they are not yet proved. However, all the 
known groundstate energies of recombined entities with few protons and electrons 
do satisfy such inequalities. } satisfied by the groundstate energies of 
Coulomb Hamiltonians $H_{N_{{\text{p}}},N_{{\text{e}}}}$ 
for $N_{{\text{p}}}+N_{{\text{e}}} \geq 3$, all the other 
contributions of more complex recombined entities decay faster 
than $\exp(\beta E_{{\text{H}}})$. For instance, the molecular 
contribution embedded in term $(N_{{\text{p}}}=2,N_{{\text{e}}}=2)$
is of order $\exp (4\beta E_{{\text{H}}})\exp(-\beta E_{{\text{H}_2}})$, 
and it decays faster than $\exp(\beta E_{{\text{H}}})$ by virtue 
of inequality~\footnote{Of course, inequality (\ref{inequalitymol}) 
also intervenes in the determination of the chemical composition of 
hydrogen \textsl{via} the minimization procedure described 
in Section~\ref{ssrecombination}. It indeed ensures that atoms are the 
proeminent chemical species at sufficiently low densities for a 
given low temperature.}
\be
\la{inequalitymol}
3 E_{{\text{H}}} < E_{{\text{H}_2}} \; .
\ee

\bigskip

Once, the contributions of molecules and other complex recombined entities 
have been discarded, it remains to estimate exchange and interaction 
contributions of ionized charges and atoms. In fact,  
since density vanishes exponentially fast, typically as 
$\exp(\beta E_{{\text{H}}})$, the mean-interparticle
distance $a=(3/(4\pi\rho))^{1/3}$ grows much faster, when $T \to 0$, 
than all thermal de Broglie wavelengths  
$\lambda_{{\text{p}}},\lambda_{{\text{e}}},\lambda_{{\text{at}}}$ 
of ionized charges and atoms. 
Also, the average interactions between those entities, which 
are proportional to inverse powers of $a$, vanish exponentially fast 
when $T \to 0$, so they become negligible compared to thermal 
energy $k_BT$. Thus exchange and interaction contributions 
can be also dropped out in the scaling limit of interest. 
According to the previous simple estimations 
and rough arguments, it is not surprising that the system 
behaves as an ideal Maxwell-Boltzmann mixture of ionized protons, 
ionized electrons and atoms, in the considered scaling limit.

\bigskip

Let us define the so-called Saha regime by
low but finite temperatures $T$, and very low but finite densities 
$\rho$ of order $\exp(\beta E_{{\text{H}}})$. According 
to the previous rigorous result, hydrogen should behave as 
a partially ionized atomic gas, weakly coupled and weakly degenerate, 
including also small fractions of other complex entities, like 
molecules $\text{H}_2$ or ions $\text{H}_2^+$ and $\text{H}^-$ for instance.
As illustrated further, it is particularly interesting,
at both practical and conceptual levels, to derive  
the corresponding corrections to the ideal terms 
obtained within Saha theory. For that purpose, it is 
convenient to combine the path integral representation 
to diagrammatical tools. Notice that the
mathematical techniques involved in previous proof 
do not provide explicit expressions for the exponentially 
small corrections beyond ideal Saha formulas.

\subsection{Path integral representation}\label{sspath}

\subsubsection{Feynman-Kac formula}

\begin{figure}
\psfragscanon
\psfrag{0}[c][c]{$\mathbf{0}$}
\psfrag{a}[c][c]{$\br_{\text{a}}$}
\psfrag{b}[c][c]{$\br_{\text{b}}$}
\psfrag{omega}[c][c]{$\bm{\omega}_{{\text{a}}{\text{b}}}
(s \beta \hbar)$}
\includegraphics[width=0.6\textwidth]{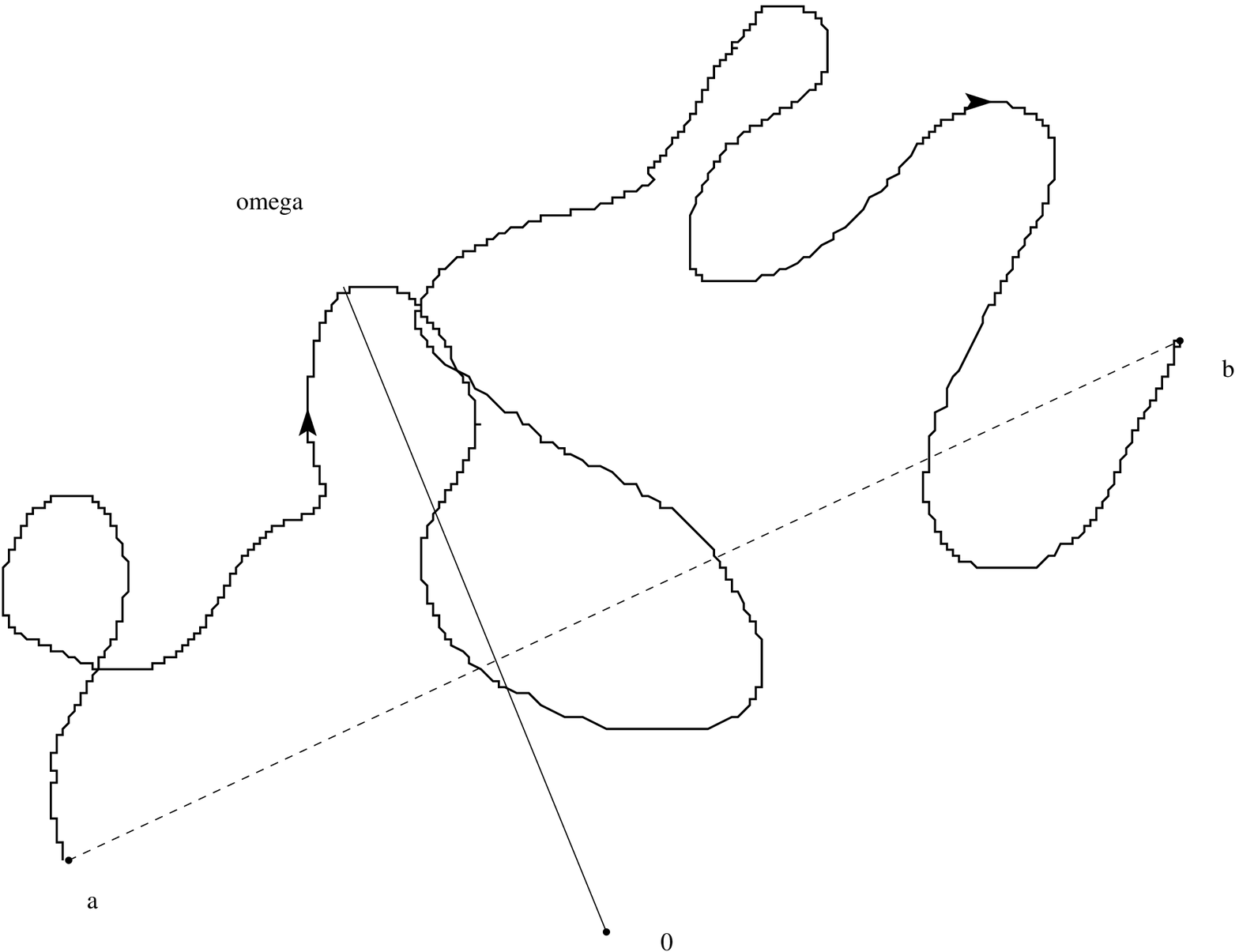}
\caption{\label{Path}A brownian 
path $\bm{\omega}_{{\text{a}}{\text{b}}}(s \beta \hbar)$. The 
dashed straight line is the uniform path connecting $\br_{{\text{a}}}$
to $\br_{{\text{b}}}$.}
\end{figure}

For the sake of pedagogy, we introduce the path integral representation in the 
simple case of a single particle with mass $m$ submitted to a 
potential $V(\br)$. Its Hamiltonian reads 
\be
\la{hamiltonian1}
H = -{\hbar^2 \over 2m} \Delta + V(\br) \; .
\ee 
The corresponding density matrix at a given temperature 
$T$, namely the matrix element of Gibbs operator $\exp(-\beta H)$,  
is exactly given by Feynman-Kac formula~\cite{Sim,Schu,Roe,Kle}
\bea
\langle \br_{{\text{b}}}|\exp (-\beta H)|\br_{{\text{a}}}  \rangle 
= \frac{\exp [-(\br_{{\text{b}}} - \br_{{\text{a}}})^2/(2\lambda^2)]}{ (2\pi \lambda^2)^{3/2}}   
\int {\cal D}(\bm{\xi}) \nonumber \\
\times \exp [-\beta \int_0^1 \dd s \; V \( (1-s)\; \br_{{\text{a}}} + s\; \br_{{\text{b}}} + 
\lambda \; \bm{\xi}(s) \) ] \; ,
\label{FK}
\eea 
with thermal de Broglie wavelength $\lambda=(\beta \hbar^2/m)^{1/2}$.
In the r.h.s. of (\ref{FK}), $\bm{\xi}(s)$ is a dimensionless 
Brownian bridge which starts from the origin at dimensionless 
time $s=0$ and comes back at the origin at dimensionless 
time $s=1$, \textsl{i.e.} $\bm{\xi}(0) = \bm{\xi}(1) = \mathbf{0}$. 
Functional measure ${\cal D}(\bm{\xi})$ is the 
normalized Gaussian Wiener measure 
which characterizes the Brownian process, and it is 
entirely defined by its covariance 
\be
\int {\cal D}(\bm{\xi})\; \xi_{{\mu}}(s)\; \xi_{{\nu}}(t) 
=\delta_{{\mu \nu}} \inf (s,t)\; (1 - \sup (s,t)) \; .
\la{covariance}
\ee
The corresponding functional integration is
performed over all Brownian bridges $\bm{\xi}(s)$. Representation
(\ref{FK}) is the proper mathematical formulation of genuine 
Feynman's idea, which amounts to express the density matrix 
as a sum over all possible paths going from $\br_{{\text{a}}}$
to $\br_{{\text{b}}}$ in a time $\beta \hbar$, of 
weighting factors $\exp (-S/\hbar)$ where $S$ is the 
classical action of a given path computed in potential $-V$. 
Here, such paths are parametrized according to
\be
\la{path}
\bm{\omega}_{{\text{a}}{\text{b}}}(s \beta \hbar)=
(1-s)\; \br_{{\text{a}}} + s\; \br_{{\text{b}}} + 
\lambda \; \bm{\xi}(s) \; ,
\ee
where $(1-s)\; \br_{{\text{a}}} + s\; \br_{{\text{b}}}$
describes the straight uniform path connecting $\br_{{\text{a}}}$
to $\br_{{\text{b}}}$ (see Fig~\ref{Path}.). Also, weighting factor 
$\exp (-S/\hbar)$ is splitted into the product of three terms. The first term, which arises 
from the kinetic energy of the straight uniform path, reduces to 
the Gaussian prefactor in front of the functional integral. The second term, 
associated with the kinetic contribution of the Brownian part of the path, 
is a Gaussian functional of $\bm{\xi}(s)$ embedded in Wiener measure 
${\cal D}(\bm{\xi})$. The third and last term is rewritten
as the Boltzmann-like factor associated with time average
\be
\la{potaverage}
\int_0^1 \dd s \; V \( \bm{\omega}_{{\text{a}}{\text{b}}}
(s \beta \hbar)  \)
\ee
of potential $V$ along the considered path 
$\bm{\omega}_{{\text{a}}{\text{b}}}$. We stress that, independently of the rather poetic introduction of path integrals by Feynman~\cite{FeyHib}, representation 
(\ref{FK}) can be derived in a straightforward way by starting from the obvious 
identity $\exp(-\beta H)=[\exp(-\beta H/N)]^N$ combined with a suitable insertion of $(N-1)$ 
closure relations in position-space (see \textsl{e.g.} 
Ref.~\cite{AlaMagPuj}). Feynman-Kac formula (\ref{FK}) then 
follows by taking the limit $N \to \infty$, as it has been proved 
for a wide class of potentials~\cite{Sim}.

\bigskip

Feynman-Kac (FK) representation (\ref{FK}) perfectly illustrates the 
intrusion of dynamical features in equilibrium static quantities 
for quantum systems, which has been quoted in Section~\ref{ssquantumscreening}. 
Let us consider the diagonal density matrix 
$\langle \br_{{\text{a}}}|\exp (-\beta H)|\br_{{\text{a}}}\rangle $.
Because of the non-commutativity of the kinetic and potential parts 
of $H$, that matrix element does not reduce to its classical 
counterpart
\be
\la{DMclassical}
\frac{\exp (-\beta  V( \br_{{\text{a}}}))}{(2\pi \lambda^2)^{3/2}} \; ,
\ee
so it is not entirely determined by $ V( \br_{{\text{a}}})$.
In fact, according to formula (\ref{FK}) specified 
to $\br_{{\text{b}}}=\br_{{\text{a}}}$, 
$\langle \br_{{\text{a}}}|\exp (-\beta H)|\br_{{\text{a}}}\rangle $
now depends on the potential landscape in some neigbourhood of 
$\br_{{\text{a}}}$ with size $\lambda$, which is explored in
time-average (\ref{potaverage}) thanks to Brownian motion. 
Therefore, $\langle \br_{{\text{a}}}|\exp (-\beta H)|\br_{{\text{a}}}\rangle $
appears to be indeed generated by a dynamical process. Not surprisingly, 
particle mass $m$ controls the importance of the 
corresponding dynamical effects. In particular, in the limit 
of an infinitely heavy particle $m \to \infty$, 
$\lambda=(\beta \hbar^2/m)^{1/2}$ vanishes and 
$\langle \br_{{\text{a}}}|\exp (-\beta H)|\br_{{\text{a}}}\rangle $
obviously tends to its classical counterpart 
(\ref{DMclassical}) : dynamical effects do not intervene anymore 
in the potential contribution which takes its purely 
static form~\footnote{Quantum corrections 
to classical formula (\ref{DMclassical}) 
can be expanded in powers of $\hbar^2$ according to 
the well-known Wigner-Kirkwood expansion~\cite{W,K,WK}. That expansion
can be easily retrieved within FK representation (\ref{FK}), 
by expanding time-average (\ref{potaverage}) in power series of 
$\lambda \bm{\xi}$ and by applying Wick theorem to 
the calculation of the resulting moments of $\bm{\xi}$.}.
Notice that the present dynamical considerations 
are the manifestation, in the framework of path integrals, of
the Heisenberg uncertainty principle which prevents the particle 
to stay at $ \br_{{\text{a}}}$. Hence, Brownian paths can be 
interpreted as describing intrinsic quantum fluctuations 
of position. 

\bigskip

Remarkably, FK representation (\ref{FK}) involves only classical objects 
and c-numbers, so the operatorial 
structure of quantum mechanics is, in some sense, erased.  
That feature turns out to be particularly useful in the 
framework of the many-body problem, as described further. However, 
the intrinsic complexity of quantum mechanics is now hidden 
in the functional integration over all Brownian bridges, which remains 
a formidable task. In fact, explicit calculations can 
be performed in a few number of cases, as reviewed in Ref.~\cite{Kle}. 
Also, aymptotic Wigner-Kirkwood expansions of 
$\langle \br_{{\text{a}}}|\exp (-\beta H)|\br_{{\text{a}}}\rangle $
around the classical formula (\ref{DMclassical}),
can be derived for situations where $\lambda$ becomes small
compared to the characteristic length $[|\na V( \br_{{\text{a}}})|/
|V( \br_{{\text{a}}})|]^{-1}$ of variation of $V(\br)$ in the 
neighbourhood of $\br_{{\text{a}}}$. Such situations 
occur for very heavy particles or for high temperatures, 
and also at large distances for 
potentials which decay as power laws at infinity. In the opposite limit
where $\lambda$ diverges, direct estimations of FK functional 
integral become rather cumbersome. Notice that, if $H$ has a single isolated 
boundstate with energy $E_0$ and wavefunction $\psi_0$, in
the zero-temperature limit, the asymptotic behaviour of  
$\langle \br_{{\text{a}}}|\exp (-\beta H)|\br_{{\text{a}}}\rangle $
is merely extracted from its spectral representation, \textsl{i.e.}
\be
\la{lowT}
\langle \br_{{\text{a}}}|\exp (-\beta H)|\br_{{\text{a}}}\rangle 
\sim |\psi_0(\br_{{\text{a}}})|^2 \exp (-\beta E_0) \;\;\;\; 
\text{when} \;\;\;\; T \to 0 \; .
\ee
As argued in Ref.~\cite{PauAlaDau}, the relevant paths which provide the 
low-temperature behaviour (\ref{lowT}) occupy a small piece of 
the whole functional phase space, and they are quite different from the 
typical paths with divergent size $\lambda$. Consequently, an exact direct estimation of their 
contribution remains an open problem in general.

\subsubsection{Gas of loops}

Let us come back now to the hydrogen plasma in the framework of the 
grand-canonical ensemble described in Section~\ref{sshydrogen}. The trace 
in the grand-partition function (\ref{gpf}) can be expressed in the 
basis of positions and spins, where a given state  
is the antisymmetrized Slater product of one-body states 
$|\mathbf{x} \; \sigma_{{\alpha}}^{{(\text{z})}} \rangle$. 
This provides a sum of 
diagonal and off-diagonal matrix elements of 
$\exp (-\beta H_{N_{{\text{p}}},N_{{\text{e}}}})$. An example of 
such matrix element with $(N_{{\text{p}}}=3,N_{{\text{e}}}=4)$ is 
\be
\la{H34} 
\langle \bR_{{1}}\bR_{{3}}\bR_{{2}}\br_{{2}}
\br_{{3}}\br_{{1}}\br_{{4}}
|\exp (-\beta H_{{3,4}})|
\bR_{{1}}\bR_{{2}}\bR_{{3}}
\br_{{1}}\br_{{2}}\br_{{3}}
\br_{{4}}  \rangle \; ,
\ee
where the positions of two protons are exchanged, 
as well as those of three electrons. Contributions of spins are 
factored out in simple degeneracy factors because the Coulomb 
Hamiltonian $H_{N_{{\text{p}}},N_{{\text{e}}}}$ does not depend on 
the spins. For matrix element (\ref{H34}), that multipliyng degeneracy 
factor is $2^4$, because the spin-states of the 
exchanged particles are necessarily identical. 

\bigskip

\begin{figure}
\psfragscanon
\psfrag{r1}[c][c]{$\br_1$}
\psfrag{r2}[c][c]{$\br_2$}
\psfrag{r3}[c][c]{$\br_3$}
\psfrag{r4}[c][c]{$\br_4$}
\psfrag{R1}[c][c]{$\bR_1$}
\psfrag{R2}[c][c]{$\bR_2$}
\psfrag{R3}[c][c]{$\bR_3$}
\psfrag{L1}[c][c]{$\cL_1$}
\psfrag{L2}[c][c]{$\cL_2$}
\psfrag{L3}[c][c]{$\cL_3$}
\psfrag{L4}[c][c]{$\cL_4$}
\includegraphics[width=0.6\textwidth]{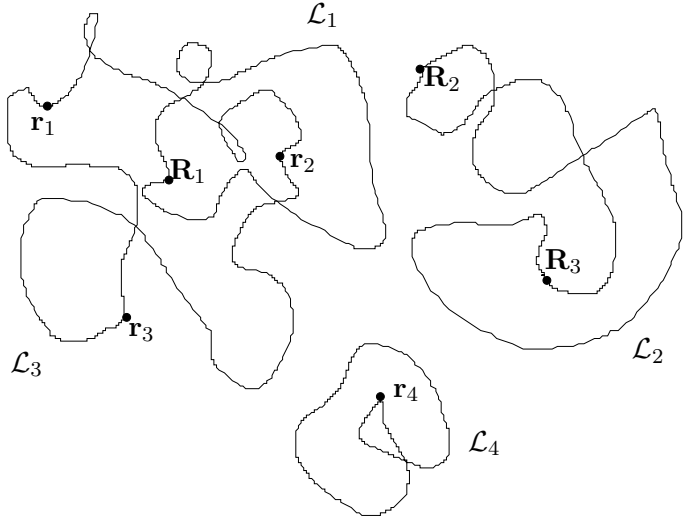}
\caption{\label{Loop}A set of four loops constructed from matrix 
element (\ref{H34}).}
\end{figure}

The FK representation for each of the matrix elements of 
$\exp (-\beta H_{N_{{\text{p}}},N_{{\text{e}}}})$ takes 
a form similar to formula (\ref{FK}), with 
$N_{{\text{p}}}$ protonic paths 
$\bm{\omega}^{{(\text{p})}}$ and $N_{{\text{e}}}$ 
electronic paths $\bm{\omega}^{{(\text{e})}}$, as well 
as a Boltzmann-like factor associated with 
the time average of the potential part
$V_{N_{{\text{p}}},N_{{\text{e}}}}$
of $H_{N_{{\text{p}}},N_{{\text{e}}}}$. The paths associated with 
matrix element (\ref{H34}) are drawn in Fig.~\ref{Loop}. 
We see that all those paths 
can be collected in loops. In fact, that 
property holds for any matrix element of 
$\exp (-\beta H_{N_{{\text{p}}},N_{{\text{e}}}})$, because 
any permutation can always be decomposed as a product of cyclic permutations. 
A loop $\cL$ is constructed by collecting $q$ paths associated with 
$q$ particles exchanged in a cyclic permutation. Accordingly, 
$\cL$ is characterized by its position $\bX$, which can be arbitrarily 
chosen among the extremities of paths $\bm{\omega}$, and several internal 
degrees of freedom which are particle species ($\alpha =\text{p},\text{e}$), 
number $q$ of exchanged particles, and shape 
$\lambda_{{\alpha}}\; \bm{\eta}$ obtained as the union of 
the $q$ paths $\bm{\omega}$. It turns out that $\bm{\eta}(s)$ is itself 
a Brownian bridge with flight time $q$, \textsl{i.e.} 
$\bm{\eta}(0)=\bm{\eta}(q)= \mathbf{0}$, distributed with the 
corresponding Wiener measure ${\cal D}(\bm{\eta})$. 

\bigskip

In the FK representation, the time-average of potential 
$V_{N_{{\text{p}}},N_{{\text{e}}}}$ can be obviously rewritten 
as a sum of two-body interactions $\cV$ between loops, plus a sum 
of loop self-energies $\cU$, namely
\be
\la{averagepotN}
\frac{1}{2} \sum_{i \neq j} \cV(\cL_i,\cL_j) + \sum_{i} \cU(\cL_i) \; ,
\ee
where loops associated with the considered matrix element are labelled 
as $\cL_i$ with index $i$ running from $1$ to their total number 
$N$. For instance, four loops can be identified in the 
FK representation of matrix element (\ref{H34}). Two-body 
potential $\cV(\cL_i,\cL_j)$ between loops $\cL_i$ and $\cL_j$ 
reduces to a time-average along 
their respective shapes of the genuine two-body particle interaction 
$v(|\bX_i + \lambda_{{\alpha_i}}\bm{\eta}_i(s)- \bX_j - \lambda_{{\alpha_j}}\bm{\eta}_j(t)|)$ 
evaluated at times which differ by an integer value. Self-energy
$\cU(\cL_i)$ for loop $\cL_i$ is given by a similar average 
along its own shape of 
$v(|\lambda_{{\alpha_i}}\bm{\eta}_i(s) - \lambda_{{\alpha_i}}\bm{\eta}_i(t)|)$ 
evaluated at times which differ by a non-zero integer value, 
with a prefactor $1/2$ which avoids double counting of 
genuine interactions between two exchanged particles.

\bigskip

At this stage, grand-partition function (\ref{gpf}) is rewritten as a 
sum of Boltzmann-like factors associated with 
energies (\ref{averagepotN}) multiplied by 
combinatorial factors and particle fugacities  
$z_{{\alpha}}/(2\pi \lambda_{{\alpha}}^2)^{3/2}$ 
with $z_{{\alpha}}= \exp (\beta \mu_{{\alpha}})$, 
which have to be integrated over positions and shapes of the 
involved loops. It turns out that the whole sum can be rewritten as the grand-partition function 
of a classical gas of undistinguishable loops with suitable 
activities $z(\cL)$, namely 
\be
\la{gpfloop}
\Xi_{\Lambda}=\Xi_{\Lambda}^{{(\text{loop})}}=
\sum_{{N=0}}^{\infty} \frac{1}{N!} \int \prod_{{i=1}}^{{N}}
\dd \cL_{{i}}\; z(\cL_{{i}}) \prod_{i<j}
\exp(-\beta \cV(\cL_i,\cL_j) ) \; .
\ee 
Phase space measure $\dd \cL$ in the world of loops, involves 
discrete summations over species index $\alpha$ and 
exchanged-particle number $q$, spatial integration over 
position $\bX$ inside $\Lambda$, and functional integration 
over shapes $\bm{\eta}$ with Wiener measure ${\cal D}(\bm{\eta})$ 
restricted to shapes such that $\bX + \lambda_{{\alpha}}\bm{\eta}(s)$
remains inside $\Lambda$. Loop fugacity reads
\be
\la{loopfug}
z(\cL)= (-1)^{q-1} \; {2z_{{\alpha}}^q \over 
q(2\pi q\lambda_{\alpha}^2)^{3/2}} \; \exp(-\beta \cU(\cL)) \; ,
\ee
the structure of which is easily intepreted as follows. 
Factor $(-1)^{q-1}$ is the signature of a cyclic permutation
of $q$ objects. In front of the obvious 
particle-activity contribution $z_{{\alpha}}^q$, factor $2$ 
is the number of configurations of exchanged-particles 
spins which are all identical. Factor $q$ in front 
of $(2\pi q\lambda_{\alpha}^2)^{3/2}$ is related to the 
$q$ possible choices of position $\bX$ among that 
of the $q$ exchanged particles. The other factor 
$q$ in front of $\lambda_{\alpha}^2$, arises from 
the absorption of the remaining $(q-1)$ integrations 
over particle positions together with the $q$ functional integrations 
over Brownian bridges into the single measure 
${\cal D}(\bm{\eta})$. Notice that $q\lambda_{\alpha}^2$ is 
nothing but the square of de Broglie wavelength 
for inverse temperature $q\beta$. 

\bigskip 

We stress that identity (\ref{gpfloop}) proceeds from  
remarkable combinatorial properties. For instance, 
in $\Xi_{\Lambda}^{{(\text{loop})}}$, there 
are various contributions in term $N=4$, which 
are identical to that of matrix element 
(\ref{H34}) in the Slater expansion of $\Xi_{\Lambda}$. Such contributions 
arise from the explicitation of the $4$ phase-space measures
$\dd \cL_i$ ($i=1,2,3,4$) where $2$ protonic loops and $2$ electronic
loops, carrying respectively $2$ or $1$ protons and $3$ or $1$ electrons, 
have to be chosen among $4$ labelled loops. Also, there are several 
matrix elements in the Slater expansion of $\Xi_{\Lambda}$, involving $3$ protons 
and $4$ electrons, which provide identical contributions to that 
of matrix element (\ref{H34}). Thanks to a rather fortunate arrangment between 
both counting factors, the full respective contributions in both 
$\Xi_{\Lambda}^{{(\text{loop})}}$ and $\Xi_{\Lambda}$ are 
indeed identical! According to those considerations, and also because of its 
synthetical form, identity (\ref{gpfloop}) is oftenly called the magic formula. 

\bigskip

Historically, Ginibre~\cite{Gin} was the first to introduce the notion 
of loops when studying the convergence of Mayer series for 
quantum gases with short-range forces. However, he did not write 
explicitely formula (\ref{gpfloop}), which has been derived later 
by Cornu~\cite{Cor}. Recently, Martin~\cite{Mar} proposed
an elegant and shorter derivation of that formula. Notice that 
identity (\ref{gpfloop}) is valid for any kind of two-body interactions, 
and an arbitrary number of species with Fermi or Bose 
statistics. For a bosonic species, factor $(-1)^{q-1}$ is 
merely replaced by $1$ in loop fugacity (\ref{loopfug}), while 
all the other factors are unchanged. Magic 
formula (\ref{gpfloop}) has been applied to a Bose gas with 
short-range interactions for studying 
Bose-Einstein condensation~\cite{MarPia}.

\bigskip

Of course, and as quoted in the simple case of a single particle in 
an external potential, the intrinsic difficulty of quantum mechanics 
is now hidden in the functional integrations over loop shapes, 
so an exact calculation of loop grand-partition function remains far beyond 
human abilities...Nevertheless, magic 
formula (\ref{gpfloop}) is quite useful because standard 
tools of classical statistical mechanics can be applied, as 
well as various transformations relying on 
simple properties of classical Boltzmann factors.

\subsection{Screened cluster representation} \label{ssSCR}

From now on, we assume that the thermodynamic limit has been taken 
once for all. Since, according to Lieb and Lebowitz theorems, 
boundary effects do not intervene anymore in the TL, we 
proceed to formal calculations in the infinite system 
where some quantities are infinite. Nonetheless, 
after suitable resummations and reorganizations, such 
divergences are removed, and it is quite reasonable to 
believe that the final results are indeed relevant, 
as far as they are finite, and they 
do describe the considered bulk quantities. That strategy is adopted 
for the calculation of particle densities and particle 
correlations. Also we work in the world of loops. According to   
magic formula (\ref{gpfloop}), those particle quantities are 
merely related to their loop counterparts. For instance, 
proton density $\rho_{{\text{p}}}$ is given in terms of 
loop density $\rho(\cL_{{\text{a}}})$ by
\be
\la{rholoop}
\rho_{{\text{p}}} = \sum_{q_{{\text{a}}}=1}^{\infty}
\int {\cal D}(\bm{\eta}_{{\text{a}}})\;  q_{{\text{a}}}\; 
\rho(\cL_{{\text{a}}}) \; ,
\ee
where loop species index is $\alpha_{{\text{a}}}=\text{p}$.

\subsubsection{Mayer graphs}

\begin{figure}
\psfragscanon
\psfrag{L1}[c][c]{$\cL_1$}
\psfrag{L2}[c][c]{$\cL_2$}
\psfrag{La}[c][c]{$\cL_a$}
\psfrag{fa1}[c][c]{$f_{a1}$}
\psfrag{fa2}[c][c]{$f_{a2}$}
\includegraphics[width=0.6\textwidth]{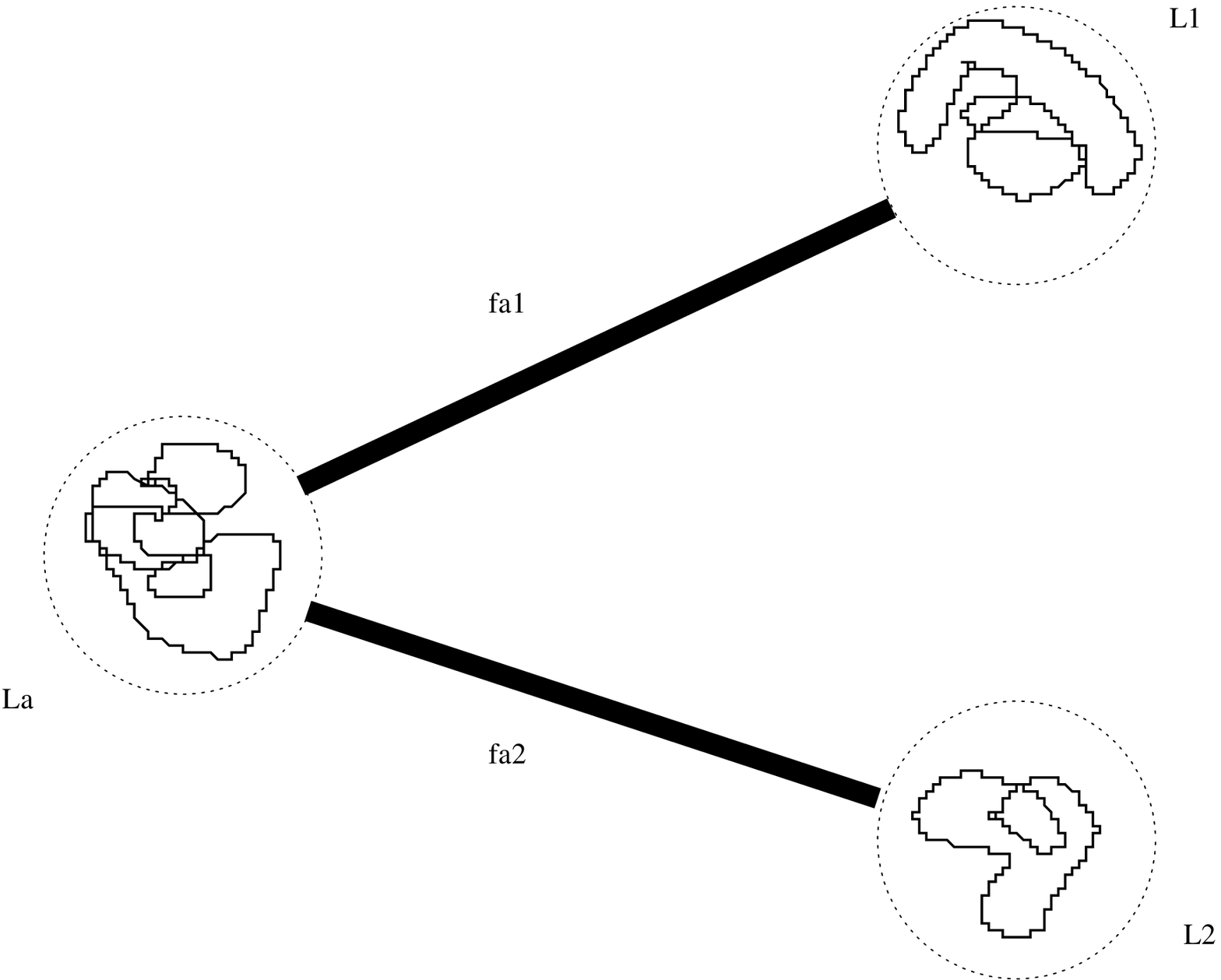}
\caption{\label{Mayer} A graph ${\cal G}$ made with root loop 
$\cL_a$ and two field loops $\cL_1$ and $\cL_2$.}
\end{figure}

The structure of loop grand-partition (\ref{gpfloop}) 
is identical to that of an ordinary classical system 
made of point particles with two-body interactions. 
Therefore, equilibrium quantities of loops can be represented 
by Mayer-like diagrammatical series, where points are replaced 
by loops. This provides Mayer series for particle quantities, 
like proton density $\rho_{{\text{p}}}$ which reads 
\be
\la{Mayerseries}
\rho_{{\text{p}}} = \sum_{\cal G} 
{1 \over S({\cal G})} \sum_{q_{{\text{a}}}=1}^{\infty}
\int {\cal D}(\bm{\eta}_{{\text{a}}}) \;  q_{{\text{a}}}
\; z(\cL_{{\text{a}}}) 
\int \prod_{i=1}^n \dd \cL_i \; z(\cL_i)
\left[ \prod f \right]_{\cal G}\; .
\ee
Each graph ${\cal G}$ is constructed 
according to the standard Mayer rules~\cite{May,Maylivre}. 
It is made with $(n+1)$ loops $\cL_i$, $i=0,...,n$ 
and $\cL_0=\cL_{{\text{a}}}$. Two loops $i$ and $j$ are connected 
at most by a single bond 
\be
\la{bond}
f_{ij}=\exp(-\beta \cV(\cL_i,\cL_j) )  - 1 \; ,
\ee
and $\left[ \prod f \right]_{\cal G}$ denotes the product of such bonds. Also, 
graph ${\cal G}$ is simply connected, namely it cannot be 
separated into two parts which are not connected by at least 
one bond $f$. Symmetry factor $S({\cal G})$ is the number 
of permutations of black loops $\cL_i$ with $i \geq 1$, which 
leave $\prod_{\cal G} f$ unchanged.
Each loop is weighted by its fugacity $z(\cL_i)$. The contribution of 
graph ${\cal G}$ is obtained by integrating over all degrees of freedom 
of black loops embedded in measure $\dd \cL_i$, while the position and 
species index of root loop $\cL_{{\text{a}}}$ are fixed and only its 
shape and particle number are integrated over. Eventually, $\sum_{\cal G}$ 
is performed over all topologically different unlabelled graphs ${\cal G}$. 
An example of graph ${\cal G}$ is shown in Fig.~\ref{Mayer}, while its contribution 
reads 
\be
\la{graph}
{1 \over 2} \sum_{q_{{\text{a}}}=1}^{\infty}
\int {\cal D}(\bm{\eta}_{{\text{a}}}) \;  q_{{\text{a}}}
\; z(\cL_{{\text{a}}}) 
\int  \dd \cL_1 \dd \cL_2 \; z(\cL_1) z(\cL_2) 
f_{{\text{a}}1} f_{{\text{a}}2}\; .
\ee

\bigskip

The loop-loop interaction is long ranged, like the genuine 
Coulomb potential itself, as illustrated by the large distance
behaviour
\be
\la{largedistance}
\cV(\cL_i,\cL_j) \sim \frac{q_ie_{{\alpha_i}}
q_je_{{\alpha_j}}}{|\bX_i-\bX_j|} \;\;\;\; 
\text{when} \;\;\;\; |\bX_i-\bX_j| \to \infty \; .
\ee 
Indeed, at large distances, loops can be shrinked to 
point charges, and further corrections to the 
monopolar form (\ref{largedistance}) can be expanded in 
multipolar power series of $\lambda_{{\alpha_i}}/|\bX_i-\bX_j|$
and $\lambda_{{\alpha_j}}/|\bX_i-\bX_j|$. 
Like in the case of purely Coulombic interactions, 
the long-range nature of $\cV(\cL_i,\cL_j)$ induces 
divergences as explained in Section~\ref{ssexamples} for 
purely Coulombic interactions. Here, in Mayer 
series (\ref{Mayerseries}), such divergences pollute every 
graph because Mayer bonds behave as $-\beta \cV(\cL_i,\cL_j)$
at large distances. 

\bigskip

In addition to previous long-range pollution,
another drawback of Mayer series relies on the difficulty 
to identify immediately the contributions of a given 
chemical species. For instance, if we 
are interested in contribution of ion $\text{H}_2^+$ 
to series (\ref{Mayerseries}), we can pick out in contribution (\ref{graph}) of graph 
drawn in Fig.~\ref{Mayer}, the term $(q_{{\text{a}}}=1,\alpha_{{\text{a}}}=\text{p}\; ; 
q_{{\text{1}}}=1,\alpha_1=\text{p}\; ; q_{{\text{2}}}=1,\alpha_2=\text{e})$, 
which involves 2 protons and 1 electron. However, because of 
the absence of bond $f_{12}$ in that graph, a proton-electron 
interaction is missing. The full interaction between the 
two protons and the single electron is recovered by summing contributions from 
all graphs ${\cal G}$ made with three loops.

\subsubsection{Resummation machinery}

\begin{figure}
\psfragscanon
\psfrag{RM}[c][c]{\large{\textbf{Resummation Machinery}}}
\psfrag{S1}[c][c]{{Expand all Mayer bonds $f$ 
in Taylor series with respect to $(-\beta \cV)$}}
\psfrag{S2}[c][c]{{Resum all convolution chains 
of bonds $(-\beta \cV)$ connecting $\cL_i$ to $\cL_j$}} 
\psfrag{S2b}[c][c]{{in terms of an effective 
potential $e_{{\alpha_i}}e_{{\alpha_j}}\phi(\cL_i,\cL_j)$}}
\psfrag{S3}[c][c]{{Introduce Boltzmann factors 
$\exp (-\beta e_{{\alpha_i}}e_{{\alpha_j}}\phi(\cL_i,\cL_j))$
and renormalized }} 
\psfrag{S3b}[c][c]{{fugacities $z_{\phi}(\cL_i)$ \textsl{via} suitable
reexponentiations}}
\psfrag{S4}[c][c]{{Identify sets of loops
${\cal E}_{\cL} = \{ \cL_i \}$ such 
that all the effective interactions between loops}} 
\psfrag{S4b}[c][c]{{inside 
${\cal E}_{\cL}$ are present in the product of Boltzmann factors 
$\exp (-\beta e_{{\alpha_i}}e_{{\alpha_j}}\phi(\cL_i,\cL_j))$}}
\psfrag{S5}[c][c]{{Identify particle clusters 
$C(N_{{\text{p}}},N_{{\text{e}}})$ in a given
set ${\cal E}_{\cL}$}} 
\psfrag{S5b}[c][c]{{by expliciting 
$\dd \cL_i$ as sums over species index and particle numbers}}
\psfrag{S6}[c][c]{{Introduce statistical weights 
$Z_{\phi}^T(C)$ for particle clusters and bonds $F_{\phi}$}} 
\psfrag{S6b}[c][c]{{\textsl{via}
suitable truncations of Boltzmann factors 
$\exp (-\beta e_{{\alpha_i}}e_{{\alpha_j}}\phi(\cL_i,\cL_j))$}}
\includegraphics[width=0.9\textwidth]{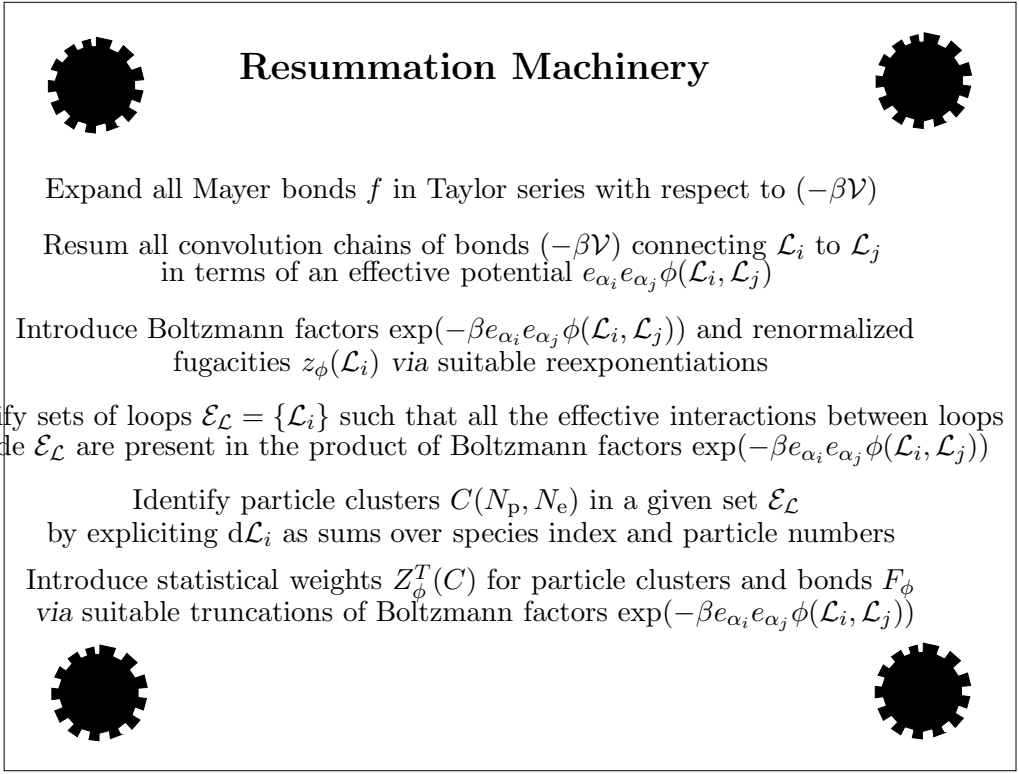}
\caption{\label{Machinery} Main steps of the resummatiom machinery.}
\end{figure}

\begin{figure}
\psfragscanon
\psfrag{Ca}[c][c]{$C_a(1,1)$}
\psfrag{C1}[c][c]{$C_1(1,1)$}
\psfrag{F}[c][c]{\Large{$\frac{\beta^2 \Phi^2}{2}$}}
\psfrag{p}[c][c]{\textbf{p}}
\psfrag{e}[c][c]{\textbf{e}}
\includegraphics[width=0.6\textwidth]{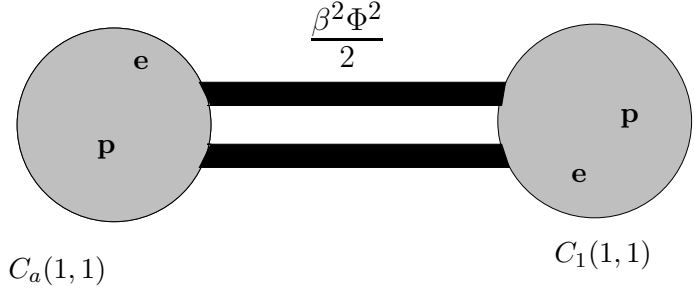}
\caption{\label{SCR} A graph $G$ made with root cluster
$C_a(1,1)$ and one field cluster $C_1(1,1)$. Each cluster contains one 
proton and one electron denoted by symbols \textbf{p} 
and \textbf{e} respectively.}
\end{figure}

The resummation machinery amounts to perform an exact transformation of 
the whole series (\ref{Mayerseries}), which takes care of drawbacks of Mayer 
grahs quoted above. The main steps are sketched in Fig.~\ref{Machinery}. 
Of course, all the transformations involve tricky counting 
calculations based on combinatorial identities, which are detailed in 
Ref.~\cite{Ala03}. The final result can be expressed as the Screened 
Cluster Representation (SCR), 
\be
\la{SCRseries}
\rho_{{\text{p}}} = \sum_{G} 
{1 \over S(G)}
\int \dd C_{{\text{a}}} \;  q_{{\text{a}}} \;
Z_{\phi}^T( C_{{\text{a}}})  
\int \prod_{i=1}^n \dd C_i \; Z_{\phi}^T(C_i)
\left[ \prod {\cal F}_{\phi}\right]_{G} \; ,
\ee 
where graphs $G$ have the same topological structure as usual Mayer graphs.  
Now, ordinary points are replaced by particle clusters $C_i$, 
weighted by factors $Z_{\phi}^T(C_i)$. Two clusters are 
connected by at most one bond $F_{\phi}$ which can be either 
$-\beta \Phi$, $\beta^2 \Phi^2/2!$ or $-\beta^3 \Phi^3/3!$ with 
$\Phi$ the total effective potential between both clusters. Graphs $G$
are simply connected, and symmetry factor $S(G)$ is computed as usual.
Phase-space measure $\dd C_i$ reduces to integrations 
over positions and shapes of loops belonging to set ${\cal E}_{\cL}$ 
associated with cluster $C_i$, whereas in $\dd C_{{\text{a}}}$ 
position of root proton is not integrated over.   
Eventually, $\sum_{G}$ is performed over all topologically different 
graphs with some simple restrictions, in part avoiding double-counting 
of genuine particle interactions~\cite{Ala03}. In Fig.~\ref{SCR}, we give 
an example of graph $G$, made with two clusters 
$C_{{\text{a}}}=C_{{\text{a}}}(1,1)$ and $C_1=C_1(1,1)$.
Its contribution reads 
\be
\la{graphG}
\int \dd C_{{\text{a}}} \;  Z_{\phi}^T( C_{{\text{a}}})  
\int \dd C_1 \; Z_{\phi}^T(C_1)
\; \frac{\beta^2 \Phi^2(C_{{\text{a}}},C_1)}{2} \; ,
\ee 
since $S(G)=1$ and $q_{{\text{a}}}=1$. Notice that part of 
graph ${\cal G}$ shown in Fig.~\ref{Mayer} contributes to previous graph 
$G$ under the action of the resummation machinery.

\bigskip

Remarkably, not only the genuine structure of Mayer graphs 
is conserved through the resummation machinery, but 
both statistical weights $Z_{\phi}^T(C)$ and bonds $F_{\phi}$
depend on the sole effective potential $\phi$. Thanks to the 
sufficient fast decay of $\phi$, all large-distances divergences 
are removed in every graph $G$, which does provide a finite contribution.
The presence of complete interactions inside a given cluster 
$C(N_{{\text{p}}},N_{{\text{e}}})$ ensures that chemical 
species made with $N_{{\text{p}}}$ protons and $N_{{\text{e}}}$
electrons do emerge at low densities. Both features are 
detailed below. Notice that the physical ideas 
underlying the SCR, are close to those involved in the construction of the 
so-called ACTEX approach by Rogers~\cite{Rog}. Also, contrarily to 
Feynman graphs involved in standard many-body 
perturbation theory~\cite{FetWal}, graphs $G$
in SCR account for non-perturbative effects in the Coulomb potential 
which are essential for a proper description of recombination.

\subsubsection{Screened effective potential}
 
Effective potential $e_{{\alpha_{\text{a}}}}e_{{\alpha_{\text{b}}}} 
\phi(\cL_{{\text{a}}},\cL_{{\text{b}}})$ 
is defined by the sum of all convolution chains 
of bonds $(-\beta \cV)$ connecting 
$\cL_{{\text{a}}}$ to $\cL_{{\text{b}}}$, namely
\be
\la{chain}
e_{{\alpha_{\text{a}}}}e_{{\alpha_{\text{b}}}} 
\phi(\cL_{{\text{a}}},\cL_{{\text{b}}}) =
\cV(\cL_{{\text{a}}},\cL_{{\text{b}}}) +
\sum_{n=1}^{\infty} (-1)^n \beta^n \int \prod_{i=1}^n \dd \cL_i \; z(\cL_i) \;
\cV(\cL_{{i-1}},\cL_{{i}}) \; \cV(\cL_n,\cL_{{\text{b}}}) \; ,
\ee
with $\cL_0=\cL_{{\text{a}}}$. 
That potential accounts for collective effects, because it involves 
contributions of an infinite number of loops. Interestingly, chain sum 
(\ref{chain}) can be reinterpreted as the complete 
perturbative expansion of the solution of integral equation 
\be
\la{intequation}
e_{{\alpha_{\text{a}}}}e_{{\alpha_{\text{b}}}} 
\phi(\cL_{{\text{a}}},\cL_{{\text{b}}})=
\cV(\cL_{{\text{a}}},\cL_{{\text{b}}}) -\beta \int \dd \cL \; z(\cL) \;
e_{{\alpha_{\text{a}}}}e_{{\alpha}} 
\phi(\cL_{{\text{a}}},\cL)
\cV(\cL,\cL_{{\text{b}}}) \; .
\ee
Therefore, $e_{{\alpha_{\text{a}}}}e_{{\alpha_{\text{b}}}} 
\phi(\cL_{{\text{a}}},\cL_{{\text{b}}})$ 
is the total potential acting on loop $\cL_{{\text{b}}}$, 
created by loop $\cL_{{\text{a}}}$
plus a cloud of loops $\cL$ distributed with density 
$-\beta  z(\cL) \; e_{{\alpha_{\text{a}}}}e_{{\alpha}} 
\phi(\cL_{{\text{a}}},\cL)$. Its physical content is then 
quite similar to that of the Debye potential 
introduced in Section~\ref{ssdebye}. In the classical 
limit $\hbar \to 0$, loops shrink to point charges so 
equation (\ref{intequation}) is nothing but the integral version 
of linearized Poisson-Boltzmann equation (\ref{helmoltz}) with 
$\epsilon_{{\text{w}}}=1$. Thus, the effective potential 
$e_{{\alpha_{\text{a}}}}e_{{\alpha_{\text{b}}}} 
\phi(\cL_{{\text{a}}},\cL_{{\text{b}}})$ can be viewed as 
the quantum analogue of the Debye potential~\cite{BMA}.

\bigskip

Similarly to what happens in the classical case, 
collective effects embedded in the effective potential are expected to screen 
the bare loop-loop interaction. In fact, the static part 
$q_{{\text{a}}}e_{{\alpha_{\text{a}}}}
q_{{\text{b}}}e_{{\alpha_{\text{b}}}}/|\bX_{{\text{a}}}-\bX_{{\text{b}}}|$ of 
$\cV(\cL_{{\text{a}}},\cL_{{\text{b}}})$, which does not 
depend on loop shapes, is exponentially screened~\footnote{This can be checked by replacing 
all loop interactions by their static parts in chain sum (\ref{chain}). The 
corresponding integration over loop internal degrees of freedom involved 
in each $\dd \cL_i$ provides the square of some wavenumber $\kappa_{{\text{loop}}}$. 
The chain sum then becomes a geometrical 
series in $-\kappa_{{\text{loop}}}^2/k^2$ in Fourier space, 
and the resulting static part of the effective potential involves factor 
$\exp (-\kappa_{{\text{loop}}}|\bX_{{\text{a}}}-\bX_{{\text{b}}}|)$.}.
On the contrary, the multipolar parts of $\cV(\cL_{{\text{a}}},\cL_{{\text{b}}})$ 
which depend on loop shapes, are not exponentially screened. The resulting 
effective potential decays only as a dipolar interaction at large distances, 
namely~\cite{BMA}
\be
\la{dipolar}
\phi(\cL_{{\text{a}}},\cL_{{\text{b}}}) \sim 
\frac{{\cal A}_{{\alpha_{\text{a}}}{\alpha_{\text{b}}}}
(\bm{\eta}_{{\text{a}}},\bm{\eta}_{{\text{b}}})}
{|\bX_{{\text{a}}}-\bX_{{\text{b}}}|^3} \;\;\;\; 
\text{when} \;\;\;\; |\bX_{{\text{a}}}-\bX_{{\text{b}}}| \to \infty \; ,
\ee
where ${\cal A}_{{\alpha_{\text{a}}}{\alpha_{\text{b}}}}
(\bm{\eta}_{{\text{a}}},\bm{\eta}_{{\text{b}}})$
is an amplitude which depends only on loop species and shapes. Consequently, bonds $F_{\phi}=-\beta\Phi$ are at the border 
line for integrability at large distances. However, if loop shapes are first integrated over, all 
dangerous $1/R^3$-terms vanish. Thus, within that prescription for the order of integrations involved in 
measures $\dd C_i$, each graph $G$ in SCR (\ref{SCRseries}) indeed provides a finite contribution.

\bigskip

The slow algebraic decay of $\phi$ can be understood as follows. 
As mentionned in Section~\ref{sspath}, paths defining loop shapes 
account for intrinsic 
quantum fluctuations of particle positions. Such fluctuations, 
which are dynamical in origin, generate $1/R^3$-dipolar interactions.
In agreement with heuristic findings argued in Section~\ref{ssquantumscreening}, 
such dynamical interactions cannot be perfectly screened 
trough collective effects, so it remains a 
$1/R^3$-tail in the large-distance behaviour of $\phi$.

\subsubsection{Emergence of chemical species}

As quoted in Section~\ref{ssrecombination}, familiar chemical species 
may emerge, strictly speaking, in a zero-density limit only. Not 
surprisingly, in such a limit, effective potential 
$e_{{\alpha_{\text{a}}}}e_{{\alpha_{\text{b}}}} 
\phi(\cL_{{\text{a}}},\cL_{{\text{b}}})$ reduces 
to the bare loop-loop interaction $\cV(\cL_{{\text{a}}},\cL_{{\text{b}}})$
as shown in Ref.~\cite{BMA}. Let us consider the contribution of the 
simple graph $G$ made with the single root cluster 
$C_{{\text{a}}}(N_{{\text{p}}},N_{{\text{e}}})$. Functional 
integrations over loop shapes of products of Boltzmann factors 
$\exp (-\beta \cV(\cL_i,\cL_j))$ embedded in $Z_{\phi}^T( C_{{\text{a}}})$, can be expressed in terms of matrix elements 
of Gibbs operators associated with Coulomb Hamiltonians, by applying 
backwards FK formula. Then, the contribution of the considered 
graph $G$ is proportional to 
\be
\la{partitionvacuum}
Z(N_{{\text{p}}},N_{{\text{e}}}) = \frac{(2\pi 
\lambda_{{N_{{\text{p}}}N_{{\text{e}}}}}^2)^{3/2}}{\Lambda}
\text{Tr} \left[\exp (-\beta H_{N_{{\text{p}}},N_{{\text{e}}}}) -...\right]
\ee
with $\lambda_{{N_{{\text{p}}}N_{{\text{e}}}}}^2=\beta\hbar^2/
(N_{{\text{p}}}m_{{\text{p}}} + N_{{\text{e}}}m_{{\text{e}}})$. 
Terms $...$ left over in the r.h.s. of definition (\ref{partitionvacuum}) indicate 
a suitable truncation of Gibbs operator 
$\exp (-\beta H_{N_{{\text{p}}},N_{{\text{e}}}})$, which ensures the 
finiteness of the trace. Thus, we see that SCR leads to a natural definition 
of the partition function $Z(N_{{\text{p}}},N_{{\text{e}}})$
for $N_{{\text{p}}}$ protons and $N_{{\text{e}}}$ electrons in the vacuum. That partition function includes the contribution of 
possible recombined species in their groundstate, as well as 
those of thermal excitations.

\bigskip

At finite densities, $Z_{\phi}^T( C_{{\text{a}}})$ accounts for 
many-body effects on cluster $C_{{\text{a}}}$, like the broadening and 
shift of energy levels of recombined species made with 
$N_{{\text{p}}}$ protons and $N_{{\text{e}}}$ electrons. Also, 
graphs $G$ with several clusters describe interactions between such clusters 
which are screened by ionized charges. For instance graph shown in Fig.~\ref{SCR}, 
involves contributions from interactions between two atoms. In fact, all 
phenomena at work can be identified in specific graphs, as argued in 
Ref.~\cite{Ala03}.

\subsection{Asymptotic expansions in the Saha regime} \label{applications}

\subsubsection{Equation of state}

Now, we consider again the Saha regime, where hydrogen behaves as a 
partially ionized atomic gas. We start with a grand-canonical description, 
where the given thermodynamic parameters are $\mu$ and $T$, and 
common particle density $\rho=\rho_{{\text{p}}}=\rho_{{\text{e}}}$ 
is a function $\rho(\mu,T)$. Making the variable change 
 $(\mu,T) \rightarrow (\gamma,T)$ defined through parametrization 
(\ref{mu_gamma}), we see that density can also be viewed as a function of 
$\gamma$ and $T$. Then, we proceed to the expansion of 
$\rho(\gamma,T)/\rho^{\ast}$ when $T \to 0$ at fixed $\gamma$, by 
investigating the corresponding behaviour of graphs $G$ in 
SCR (\ref{SCRseries}). Roughly speaking, that behaviour results from the 
competition between three kind of contributions, which vanish or explode 
exponentially fast when $T \to 0$, namely

\bigskip 

\noindent (i) $\gamma^N \exp( \beta N E_{{\text{H}}})$ 
with $N=N_{{\text{p}}}+N_{{\text{e}}}$ arising from
entropy factor $\exp(\beta \mu N)$ in $Z_{\phi}^T(C)$ 

\bigskip

\noindent (ii) $\exp( -\beta  
E_{{N_{\text{p}},N_{\text{e}}}}^{(0)})$ arising from a recombined 
entity with groundstate energy $E_{{N_{\text{p}},N_{\text{e}}}}^{(0)}$  
in $Z_{\phi}^T(C)$

\bigskip

\noindent (iii)  $\gamma^{p/2} \exp( \beta p E_{{\text{H}}}/2)$ 
with $p$ relative integer, arising from interactions screened over 
Debye length 
$\lambda_{{\text{D}}} = (4\pi\beta e^2 (\rho_{{\text{p}}}^{{(\text{id})}}+
\rho_{{\text{e}}}^{{(\text{id})}}))^{-1/2}$ 
with $\rho_{{\text{p}}}^{{(\text{id})}}=
\rho_{{\text{e}}}^{{(\text{id})}} =
\rho^{\ast}\gamma$ in bonds ${\cal F}_{\phi}$
between charged clusters~\footnote{Since ionized protons and ionized 
electrons behave almost classically in the Saha regime, contributions of 
screened interactions are controlled by Debye length 
associated with the corresponding density of those charges. Notice 
that screened potential $e_{{\alpha_{\text{a}}}}e_{{\alpha_{\text{b}}}} 
\phi(\cL_{{\text{a}}},\cL_{{\text{b}}})$ 
precisely decays as 
$q_{{\text{a}}} e_{{\alpha_{\text{a}}}} 
q_{{\text{b}}} e_{{\alpha_{\text{b}}}} \exp (-\kappa_{\text{D}}|\bX_{{\text{a}}}-\bX_{{\text{b}}}|)
/|\bX_{{\text{a}}}-\bX_{{\text{b}}}|$ at distances 
$|\bX_{{\text{a}}}-\bX_{{\text{b}}}|$ of order 
$\lambda_{{\text{D}}}$. The algebraic
$1/|\bX_{{\text{a}}}-\bX_{{\text{b}}}|^3$-decay of $\phi$ 
takes place at distances $|\bX_{{\text{a}}}-\bX_{{\text{b}}}|$ 
much larger than $\lambda_{{\text{D}}}$~\cite{Ala08}.}

\bigskip

\noindent Previous analysis provides the Scaled Low Temperature expansion of 
$\rho(\gamma,T)/\rho^{\ast}$, where leading terms 
\be
\la{leadingrho}
\gamma + \frac{\gamma^2}{2}
\ee
are indeed those predicted by Saha theory. Each correction reduces to some 
power of $\gamma$ multiplied by a temperature-dependent function 
which decays exponentially fast when $T \to 0$, in agreement with 
the rigorous behaviour (\ref{rhoSL}) presented in Section~\ref{sshydrogen}.

\bigskip

The SLT expansion of $\beta P/\rho^{\ast}$ is readily obtained from that 
of $\rho(\gamma,T)/\rho^{\ast}$, by integrating identity
\be
\la{thermoidentity}
\frac{\partial \beta P}{\partial \gamma}(\beta,\gamma)=\frac{2\rho}{\gamma} \;
\ee
whichs follows from the standard thermodynamical relation 
between $\rho$, $P$ and $\mu$. A further elimination of $\gamma$ between 
both SLT expansions of $\beta P(\gamma,T)/\rho^{\ast}$ 
and $\rho(\gamma,T)/\rho^{\ast}$ provides the SLT expansion of the 
equation of state, \textsl{i.e.} the asymptotic 
expansion of pressure in units of $\rho^{\ast} k_{{\text{B}}}T$ 
when $T \to 0$ at fixed ratio $\rho/\rho^{\ast}$,
\be
\la{EOSSLT}
\beta P/\rho^{\ast}=\beta P_{{\text{Saha}}}/\rho^{\ast} 
+ \sum_{k=1}^{\infty} b_k(\rho/\rho^{\ast}) \alpha_k(\beta) \; .
\ee
The leading term is indeed given by Saha formula (\ref{Sahapressure}). 
Coefficients $b_k(\rho/\rho^{\ast})$ are algebraic functions of ratio
$\rho/\rho^{\ast}$, while temperature-dependent functions
$\alpha_k(\beta)$ decay exponentially fast when $T$ vanishes,
$\alpha_k(\beta) \sim \exp(-\beta \delta_k)$ except for possible
multiplicative powers of $\beta$. Expansion \eqref{EOSSLT}
is ordered with respect to increasing decay rates,
$0 < \delta_1 < \delta_2 < ...$. The first five corrections have been
computed in Ref.~\cite{Ala08} (see Ref.~\cite{AlaBal} 
for a pedagogical presentation of the calculations).
In the following table, we summarize 
their physical content, as well as the expressions and values of the corresponding decay rates which are merely obtained by taking 
the products of above exponential factors (i), (ii) and (iii).
 
\bigskip

\begin{tabular}{|c|l|r|}
\hline
Correction ($k$)	&	Physical content	& $\delta_k$ (in eV) \\ \hline
1 & plasma polarization around ionized charges & $|E_{{\text{H}}}|/2 \simeq 6.8$ \\
2 & formation of molecules, atom-atom interactions & $|3E_{{\text{H}}}-E_{{\text{H}_2}}| \simeq 9.1$ \\
3 & atomic excitations, charge-charge interactions & $3|E_{{\text{H}}}|/4 \simeq 10.2$ \\
4 & formation of ions, atom-charge interactions & $|2E_{{\text{H}}}-E_{{\text{H}_2^+}}| \simeq 11.0$ \\
5 & fluctuations of plasma polarization & $|E_{{\text{H}}}| \simeq 13.6$ \\ \hline
\end{tabular}

\bigskip

Remarkably, all non-ideal corrections to Saha equation of state 
evocated in Section~\ref{sshydrogen} are properly identified and ordered 
in exact SLT expansion (\ref{EOSSLT}). Contributions of usual chemical 
species do emerge in the present double low-temperature and low-density limit, 
in agreement with heuristic findings presented in 
Section~\ref{ssrecombination}. Their relative importance, determined by 
the ordering of decay rates $\delta_k$, results from subtle 
inequalities between groundstate 
energies $E_{{N_{\text{p}},N_{\text{e}}}}^{(0)}$ and 
$E_{{\text{H}}}$ in the vacuum. For instance, formation of 
molecules $\text{H}_2$ dominate that of ions $\text{H}_2^+$ or 
$\text{H}^-$, while contributions 
of more complex entitites, like $\text{H}_2^-$, $\text{H}_3^+$ or
$\text{H}_3$, decay exponentially faster than 
$\exp( -\beta | E_{{\text{H}}}|)$ 
as detailed in Ref.~\cite{Ala08}. Thermal excitations of atoms, 
molecules and ions are accounted for in functions $\alpha_k(\beta)$ 
\textsl{via} the corresponding cluster partition functions 
$Z(1,1)$, $Z(2,2)$, $Z(2,1)$ and $Z(1,2)$. However, we stress that 
such contributions of recombined entities are entangled to that 
of their dissociation products. For instance, in $Z(1,1)$, contributions 
from excited atomic states cannot be separated from that of ionized 
protons and ionized electrons. Also, those cluster partition functions, 
built with bare Coulomb Hamiltonians,    
are indeed finite thanks to the substraction of non-integrable 
long-range parts, which are ultimately screened by ionized charges. 
Thus, only the full contribution of those species partition functions 
and of their screened long-range parts, makes physical sense. For instance, 
term $k=3$ involves both contributions of $Z(1,1)$ and of screened proton-electron interactions. In that context, the extraction from 
$\alpha_3(\beta)$ of a purely atomic contribution, for instance given by
the phenomenological expression (\ref{internalpartitionbis}) 
of $Z_{{\text{H}}}^{{(\text{scr})}}$ or by the Planck-Larkin 
formula~\cite{PlaLar}, remains arbitrary. We stress that 
such arbitrariness does not cause any trouble here, since
only the full contribution embedded in $\alpha_3(\beta)$ 
is relevant for thermodynamics.

\bigskip

Eventually, notice that ordinary virial expansions in powers of $\rho$ 
at fixed $T$~\cite{Ebe,Kra,AlaPer,Kha}, 
can be easily recovered from SLT expansion (\ref{EOSSLT}) by expanding 
coefficients $b_k(\rho/\rho^{\ast})$ in powers of $\rho/\rho^{\ast}$.
This has been explicitely checked up to order $\rho^2$, by noting that 
partition functions $Z(1,1)$, $Z(2,0)$ and $Z(0,2)$ are 
merely related to two-body quantum virial functions first introduced 
by Ebeling~\cite{Ebe}.

\subsubsection{Particle correlations}

The SCR also provides diagrammatical series for equilibrium 
particle correlations, 
$\rho_{{\alpha_{\text{a}}\alpha_{\text{b}}}}^{({\text{T}})}
(\br_{{\text{a}}},\br_{{\text{b}}})=
\rho_{{\alpha_{\text{a}}\alpha_{\text{b}}}}^{({2})}
(\br_{{\text{a}}},\br_{{\text{b}}})-
\rho_{{\alpha_{\text{a}}}}\rho_{{\alpha_{\text{b}}}}$, 
where graphs are similar to those 
introduced above for particle densities~\cite{Ala03}. Not surprisingly, 
the asymptotical behaviour of $\rho_{{\alpha_{\text{a}}\alpha_{\text{b}}}}^{({\text{T}})}
(\br_{{\text{a}}},\br_{{\text{b}}})$ 
when $R=|\br_{{\text{a}}}-\br_{{\text{b}}}| \to \infty$, 
is determined by the large-distance behaviour of bonds ${\cal F}_{\phi}$.
It turns out that bonds $-\beta \Phi$, which might give 
\textsl{a priori} $1/R^3$-contributions, ultimately provide short-range 
terms thanks to the rotational invariance of statistical weights of 
particle clusters combined to the harmonicity of Coulomb 
potential~\cite{AlaCorMar}. On the contrary, bonds   
$\beta^2 \Phi^2/2!$ provide $1/R^6$-contributions which do not cancel out. 
Thus, according to that graph by graph analysis~\cite{AlaCorMar}, 
all particle correlations are expected to decay as $1/R^6$ when 
$R \to \infty$, namely
\be
\la{algebraictail}
\rho_{{\alpha_{\text{a}}\alpha_{\text{b}}}}^{({\text{T}})}
(\br_{{\text{a}}},\br_{{\text{b}}}) \sim 
\frac{A_{{\alpha_{\text{a}}\alpha_{\text{b}}}}(\beta,\rho)}{R^6}\; ,
\ee
whith temperature- and density-dependent amplitudes 
$A_{{\alpha_{\text{a}}\alpha_{\text{b}}}}(\beta,\rho)$.

\bigskip

In the Saha regime, amplitudes $A_{{\alpha_{\text{a}}\alpha_{\text{b}}}}(\beta,\rho)$ 
can be determined within the method used for deriving the SLT 
expansion of the pressure (\ref{EOSSLT}). Here, it is important to select 
first the graphs which contribute to the $1/R^6$-tails, and 
afterwards to take the scaled low-density and low-temperature limit 
of the corresponding contributions. That order of limits ensures 
that collective screening effects arising from ionized 
protons and ionized electrons are indeed taken into account. Notice 
that Debye screening length $\lambda_{{\text{D}}}$ 
associated with those almost classical 
ionized charges diverges in the SLT limit.

\bigskip
 
Within above procedure, 
SLT expansions of $A_{{\alpha_{\text{a}}\alpha_{\text{b}}}}(\beta,\rho)$ 
can be derived~\cite{AlaCorMar}. At leading order, 
$A_{{\text{p}\text{p}}}(\beta,\rho)$ reduces to a quadratic form 
in the ideal densities (\ref{rhoideal1}) of (\ref{rhoideal2}) of 
ionized protons and hydrogen atoms respectively,
\be
\la{quadratic}
A_{{\text{p}\text{p}}}(\beta,\rho)= \rho_{{\text{at}}}^{{(\text{id})}}
\rho_{{\text{at}}}^{{(\text{id})}}
C_{{\text{at}-\text{at}}}(T)
+ \rho_{{\text{p}}}^{{(\text{id})}}
\rho_{{\text{at}}}^{{(\text{id})}}
C_{{\text{p}-\text{at}}}(T)
+\rho_{{\text{at}}}^{{(\text{id})}}
\rho_{{\text{p}}}^{{(\text{id})}}
C_{{\text{at}-\text{p}}}(T) +...
\ee
where terms left over decay exponentially faster than 
$(\rho^{\ast})^2$. That structure is 
easily interpreted by noting that each proton in correlation 
$\rho_{{\text{p}\text{p}}}^{({\text{T}})}$ may be either ionized
or recombined into an atom. Then, the insertion of that leading quadratic 
structure of amplitudes in asymptotical behaviour (\ref{algebraictail})
provides a natural definition of 
effective potentials between ionized charges and atoms. In particular, 
the atom-atom effective interaction is 
\be
\la{atomatomeff}
U_{{\text{at}-\text{at}}}^{({\text{eff}})}(R) =
-\frac{k_{{\text{B}}}T \; C_{{\text{at}-\text{at}}}(T)}{R^6} \; .
\ee
Taking into account the expression of $C_{{\text{at}-\text{at}}}(T)$
derived in Ref.~\cite{AlaCorMar}, and definition (\ref{vdWpotential}) 
of van der Waals potential, that effective interaction can be rewritten 
as 
\be
\la{effectivevdW}
U_{{\text{at}-\text{at}}}^{({\text{eff}})}(R)=
\(1-c_{{\text{at}}}
\frac{k_{{\text{B}}}T}{|E_{{\text{H}}}|} \) \; 
U_{{\text{vdW}}}(R)
\ee
where $c_{{\text{at}}}$ is a numerical positive constant. Thus, 
atom-atom van der Waals interactions are not perfectly screened by free 
charges, which only reduce their amplitude, in qualitative 
agreement with heuristic arguments presented 
in Section~\ref{ssrecombination}. Remarkably, reduction factor 
$(1-c_{{\text{at}}}
k_{{\text{B}}}T/|E_{{\text{H}}}|)$ does not depend on 
the proportion of free charges, since positive constant 
$c_{{\text{at}}}$ can be expressed in terms of 
the sole atomic spectrum~\cite{AlaCorMar}. The density of free 
charges only intervenes in Debye screening length 
$\lambda_{{\text{D}}}$, while formula (\ref{effectivevdW}) 
holds for $R \gg \lambda_{{\text{D}}}$. Notice that, 
in the window $a_{{\text{B}}} \ll R \ll \lambda_{{\text{D}}}$,
$U_{{\text{at}-\text{at}}}^{({\text{eff}})}(R)$ does 
reduce to $U_{{\text{vdW}}}(R)$ apart from exponentially 
small corrections with $T$ arising from atomic excited 
states~\cite{AlaCorMar}.

\subsection{Conclusion} \label{conclusion}

According to the exact results described here, and also to other 
works in the literature, the present state of the art for 
the various issues about screening, 
recombination and van der Waals forces is summarized below, 
together with some comments about related open problems. 
Part of those results are described in Ref.~\cite{BryMarreview} 
where exact results for Coulomb systems at low density are reviewed.

\bigskip

\noindent $\bullet$ \textsl{ Debye 
exponential screening is destroyed by quantum fluctuations}

\bigskip

\noindent Quantum fluctuations of positions 
generate instantaneous electrical dipoles which cannot be 
perfectly screened by the surrounding plasma because of their 
dynamical character. Fluctuations of the resulting $1/R^3$-dipolar effective 
interactions ultimately pollute equilibrium particle correlations 
with $1/R^6$-algebraic tails. Such fluctuations are not taken into 
account in usual mean-field theories, like Thomas-Fermi or RPA, which 
erroneously predict an exponential decay~\cite{AlaIUPAP,CorMar}. 
Previous mechanism, intrinsic to 
quantum mechanics, is present in any thermodynamical state. Explicit 
perturbative calculations of $1/R^6$-tails have been performed in regimes 
where free charges behave almost 
classically~\cite{AlaCorMar,CorTail}, for which 
such tails appear at distances $R \gg \lambda_{{\text{D}}}$. 
Also, it has been shown~\cite{AlaMar89}, through a non-perturbative analysis, 
that the effective potential between two quantum charges 
immersed in a classical plasma does decay as $1/R^6$. Thus, all those results 
strongly suggest the breakdown of exponential screening, 
although a rigorous derivation is not yet available.

\bigskip

Besides a proof of algebraic screening, explicit calculations of 
algebraic tails in strongly degenerate plasmas remain to be done. 
In particular, such calculations might be quite useful in 
condensed matter, where effective electron-electron interactions 
are oftenly modelized as Dirac delta functions in order to account 
for screening~\footnote{Such a modelization of screening of 
Coulomb interactions between electrons, may be sufficient for 
interpreting rather complicated phenomena mainly driven by 
other mechanisms. Nevertheless, the existence of long-ranged normal
attractive interactions between electrons might 
induce some observable consequences. For instance, in normal
superconductivity, interactions generated \textsl{via} the 
electron-phonon coupling are believed to be responsible for the 
formation of Cooper pairs.}.

\bigskip

\noindent $\bullet$ \textsl{Recombined entities must be defined in a 
double zero-density and zero-temperature limit}

\bigskip

\noindent In agreement with simple findings, 
contributions to thermodynamical quantities of 
familiar chemical species, emerge unambiguously when both 
density and temperature vanish. The Screened Cluster Representation, 
which can be devised for any mixture of nuclei and 
electrons~\cite{Ala03}, is a suitable 
theoretical framework for evaluating such contributions. As it 
can be naively expected,  when $T \to 0$, the leading contribution of 
a given recombined entity is controlled by the Boltzmann factor 
associated with its groundstate energy in the vacuum. At finite temperatures,
contributions from thermal excitations of that entity, and from its 
dissociation products and their screened interactions, 
are all mixed together.
This is well illustrated by the analysis of the equation of state 
of hydrogen in the Saha regime~\cite{Ala08}.

\bigskip

In approaches based on the chemical picture, the choice of 
a suitable internal partition function for recombined species 
is a central question, which has
been the source of many controversies since the
introduction of Planck-Larkin formula
(see e.g. Refs.~\cite{Ios,Nor,Sta}). As far as 
thermodynamical quantities are concerned, the analysis 
of the various contributions in their SCR confirms 
the arbitrariness of such a choice. More interesting, that analysis 
shows that, in phenomenological theories, the internal partition function
of a given chemical species, 
should be defined simultaneously with the 
related contributions of its elementary components. The SCR 
itself might serve as an useful guide for introducing 
suitable modelized ingredients in chemical approaches. 

\bigskip

\noindent $\bullet$ \textsl{Free charges reduce the amplitude of van der Waals interactions}

\bigskip

\noindent In the framework of the many-body problem, it is natural to define 
effective potentials between recombined entities from the asymptotic large-distance behaviour 
of equilibrium correlations between their elementary components. 
Since all particle correlations decay as $1/R^6$, effective interactions between 
neutral or charged entities also decay \textsl{\`a la van der Waals} as $1/R^6$.
The corresponding amplitudes are controlled by quantum fluctuations of 
electrical dipoles, the size of which is typically of order either the Bohr radius
$a_{{\text{B}}}$ for recombined charges or thermal de Broglie wavelengths 
for free charges. Thus, genuine van der Waals interactions between atoms or 
molecules are not perfectly screened by free charges, as 
a consequence of the breakdown of Debye exponential screening.

\bigskip

Free charges should reduce the amplitude of van der Waals interactions,
as suggested by the exact calculation for hydrogen in the Saha regime. 
At sufficiently low temperatures and low densities, the corresponding 
renormalization factor does not depend on $\rho$, and is linear in $T$. 
It should be quite instructive to compare that first-principles prediction 
to that of phenomenological 
approaches like Lifchitz theory~\cite{Lif,LifLan,Nin,Par}. Also, the SCR of 
particle correlations should provide some insights for 
reliable estimations of previous reduction factor at higher 
temperatures and densities. The reduction of the amplitude 
of van der Waals interactions by free charges might have important 
consequences on the phase diagrams, as quoted in Ref.\cite{Paral} 
for a system of biological macromolecules immersed in an highly 
concentrated salt.

\end{document}